\documentclass[%
 reprint,
superscriptaddress,
 amsmath,amssymb,
 aps,
]{revtex4-1}

\usepackage{graphicx}
\usepackage{dcolumn}
\usepackage{bm}
\usepackage{color}
\usepackage[mathlines]{lineno}
\usepackage{soul}

\newcommand{\be}{\begin{equation}}
\newcommand{\ee}{\end{equation}}

\begin{document}

\preprint{APS/123-QED}

\title{Deep Learning the Morphology of Dark Matter Substructure}

\author{Stephon Alexander}

 \affiliation{Brown Theoretical Physics Center, Brown University, Providence, RI, USA}
 \affiliation{%
 Department of Physics, Brown University, Providence, RI, USA
}%

\author{Sergei Gleyzer}

\affiliation{
 Department of Physics and Astronomy, University of Alabama, Tuscaloosa, AL, USA 
}%

\author{Evan McDonough}%
\affiliation{Brown Theoretical Physics Center, Brown University, Providence, RI, USA}
 \affiliation{%
 Department of Physics, Brown University, Providence, RI, USA
}%

\author{Michael W. Toomey}
 \email{michael\_toomey@brown.edu}
\author{Emanuele Usai}
\affiliation{%
 Department of Physics, Brown University, Providence, RI, USA
}%

\begin{abstract}

Strong gravitational lensing is a promising probe of the substructure of dark matter halos. Deep learning methods have the potential to accurately identify images containing substructure, and differentiate WIMP dark matter from other well motivated models, including vortex substructure of dark matter condensates and superfluids. This is crucial in future efforts to identify the true nature of dark matter. We implement, for the first time, a classification approach to identifying dark matter substructure based on simulated strong lensing images with different substructure. Utilizing convolutional neural networks trained on sets of simulated images, we demonstrate the feasibility of deep neural networks to reliably distinguish among different types of dark matter substructure. With thousands of strong lensing images anticipated with the coming launch of LSST, we expect that supervised and unsupervised deep learning models will play a crucial role in determining the nature of dark matter.

\end{abstract}

\maketitle

\section{Introduction}

The canonical candidate for dark matter is a weakly interacting massive particle (WIMP). Indeed, extensions of the Standard Model (SM) generally include WIMPs of mass 100 GeV that accurately reproduce the observed dark matter density; realizing what is known as the \textit{WIMP miracle}. However, WIMPS have thus far evaded detection, both by direct detection \cite{Drukier:1986tm,Goodman:1984dc,Akerib:2016vxi,Cui:2017nnn,Aprile:2018dbl} and colliders (e.g. \cite{Aaboud:2019yqu}). There  are also hints at cracks in the WIMP paradigm,  for example, the core vs. cusp problem: observations of halos have consistently shown that actual dark matter halos lack cusps \cite{burk} like that of the Navarro–-Frenk–-White (NFW) profile found from simulation \cite{nfw}. This motivates the consideration of alternatives to the WIMP paradigm.

An interesting possibility is \emph{condensate} models of dark matter, both Bose-Einstein (BEC) \cite{Sin:1992bg,Silverman:2002qx,Hu:2000ke,Sikivie:2009qn,Hui:2016ltb,Berezhiani:2015bqa,Ferreira:2018wup} and Bardeen-Cooper-Schreifer (BCS) \cite{Alexander:2016glq,Alexander:2018fjp}. These build on the decades-long study of axion dark matter \cite{Preskill:1982cy,Abbott:1982af,Dine:1982ah} and the realization that axions, arising as the Goldstone boson of a spontaneously-broken global U(1) symmetry, are the field theory definition of superfluidity \cite{Schmitt:2014eka}. In these models, dark matter is a quasi-particle excitation of the fundamental degrees of freedom that comprise the condensate. For a specific choice of the effective field theory of the superfluid, this reproduces the baryonic Tully--Fisher relation \cite{Berezhiani:2015bqa,Berezhiani:2015pia}. 

These condensate models have the interesting property that they can form \emph{vortices} \cite{vort}, line-like defects that are a non--relativistic analog to cosmic strings \cite{Brandenberger:1993by,Brandenberger:2013tr}. If they exist, vortices constitute a substructure component for dark matter halos. The detection of vortices would be a smoking gun for superfluid dark matter. We are thus lead to discriminate between different models of dark matter by probing substructure in halos. 

In practice, the best method to detect substructure is from strong gravitational lensing images. Observations of strongly lensed quasars have been previously used to infer the presence of substructure \cite{sub1,sub2,sub3}. Additionally, high resolution images with ALMA have inferred the presence of sub-galactic structure \cite{alma}. Extended lensing images, in particular, can serve as very sensitive probes of underlying dark matter substructure \cite{veg,koop,veko}. Given strong lensing has already proven to be a powerful probe of dark matter substructure, it is logical to extend this to distinguishing between different types of dark matter substructure. 

Bayesian likelihood analyses can be implemented to determine if a given dark matter model is consistent with a set of lensing images. Indeed, such analyses have been conducted searching for particle dark matter substructure \cite{pcat,subs}. In this work we take a different approach, and with condensate models of dark matter in mind, implement a deep learning algorithm to identify specific types of dark matter in simulated lensing images; that is, we consider the search for substructure as a classification problem.

Applications of machine and deep learning methods are abound in cosmology \cite{Ntampaka:2019udw} and the physical sciences more broadly \cite{Carleo:2019ptp}. In particular, this approach has been applied to strong gravitational lensing \cite{Hezaveh:2017sht,PerreaultLevasseur:2017ltk,Morningstar:2018ase,Morningstar:2019szx}, and most recently, to the study of particle dark matter sub-halos \cite{Brehmer:2019jyt}.

The treatment of substructure searches as a \emph{classification problem} compliments the existing approaches of statistical detection \cite{Rivero:2017mao,Cyr-Racine:2015jwa,Cyr-Racine:2018htu,Rivero:2018bcd,Brennan:2018jhq} and identification of individual substructures (e.g. \cite{alma}). This work can be interpreted as an intermediate step before the latter: we train and implement a convolutional neural network (CNN) to distinguish among different classes of substructure in lensing images that can then be further processed to find the position, mass, and other properties, of individual substructures.

The overarching goal of this work is to undertake a theory-agnostic approach to dark matter searches. As a first step, we first present the results of an implementation of a supervised neural network to distinguish between two different types of dark matter. Given the vast number of models and considerable theoretical uncertainty on the nature of dark matter, it would then be advantageous to implement an \emph{unsupervised} machine learning algorithm to identify various potential dark matter signals in the strong lensing images. 

The structure of this paper is as follows: in section \ref{substuct} we review dark matter substructure, and in section \ref{superfluid} we consider as a prototypical example the substructure of superfluid dark matter. We construct simulated lensing images in section \ref{sec:stronglensing} and in section \ref{CNN} a neural network to analyse them. We present our results in section \ref{results}, and discuss the implications for detection in section \ref{detection}. We close with a discussion of future work in section \ref{discussion}. 

\section{Dark Matter Substructure and Strong Gravitational Lensing}

\label{substuct}

The $\Lambda$CDM paradigm predicts that density fluctuations present in the the early universe evolve to become the large scale structure of the universe via hierarchical structure formation. This model envisions small halos merge together forming larger and larger structures leading to the DM halos that we see today \cite{Kauffmann:1993gv}.  It is also expected from simulation that subhalos can avoid significant tidal disruption and remain largely intact. On large scales $\Lambda$CDM is consistent with the CMB, galaxy clustering, and weak lensing \cite{planck,clust,wlens}. However, on smaller scales the verdict is less clear. A classic example is the missing satellites problem \cite{msp} (though see \cite{mspr} for a differing take). Indeed, different types of particle dark matter can have vastly different substructure on subgalactic scales. For example, the greater streaming length of WDM \cite{wdm1,wdm2}  and emergent properties of self--interacting dark matter \cite{idm} can prevent the formation of small scale substructure. Thus, while large scale structure for different types of DM can appear identical, careful attention to structure on subgalactic scales can be a powerful tool to distinguish DM models.

A powerful probe of the gravitationally bound structures of dark matter is strong gravitational lensing. Given a matter over or under density, the deflection angle along the line of sight is given by an integral over the induced gravitational potential \cite{Nar-Bart:1997lens},
\begin{equation}
    \vec{\alpha} = \frac{2}{c^2} \vec{\nabla}_{\theta} \int {\rm d}\chi \frac{\chi_{s} - \chi}{\chi \chi_s} \Psi(\vec{r}) ,
\end{equation}
where $\chi$ is the distance along the line of sight, $\chi_s$ is the distance to the source, and $\Psi(\vec{r})$ is the gravitational potential. In the thin lens approximation \cite{Nar-Bart:1997lens}, valid in the limit that the thickness of the lensing galaxy is small compared to the distance to the lens, this takes a simplified form, 
\begin{equation}
    \vec{\alpha} = \frac{2}{c^2} \frac{D_{LS}}{D_S D_{L}} \vec{\nabla} \int {\rm d}z \Psi(\vec{r}) ,
\end{equation}
where $D_{LS}$, $D_L$, and $D_S$ are the angular diameter distances from the lens to the source, from the observer to the lens, and from the observer to the source, respectively, and $z$ is the distance along the line of sight.  From this expression one can straightforwardly compute the lensing due to any gravitational potential $\Psi$.

The gravitational potential is in turn determined by matter density via the Poisson equation, $\nabla^2 \Psi \propto \rho$. The linearity of this equation implies that the total lensing due to the separate contributions, e.g. of a halo and halo substructure, is simply the sum of the individual contributions. That is,
\begin{equation}
    \vec{\alpha} = \vec{\alpha}_{LSS} + \vec{\alpha}_{halo} + \vec{\alpha}_{halo-sub},
\end{equation}
where $\vec{\alpha}_{LSS}$ is the external shear due to large scale structure, and $\vec{\alpha}_{halo,halo-sub}$ are the lensing due to the halo and halo substructure respectively.  

The well studied case is the lensing due to the spherical substructures expected from hierarchical structure formation in the context of non-interacting particle dark matter. However, as mentioned in the introduction, other types of substructure can exist in models of dark matter outside the WIMP paradigm. As a prototypical example, we will consider dark matter condensates, namely superfluids, which exhibit substructure in the form of \emph{vortices}. We now proceed to develop this in detail.

\section{Case Study: Dark Matter Superfluidity}
\label{superfluid}

The canonical example of a condensate dark matter model is axion dark matter. Axions were introduced as a solution to the strong-CP problem of the standard model \cite{Peccei:1977hh,Wilczek:1977pj,Weinberg:1977ma}, and soon there-after proposed as a dark matter candidate \cite{Preskill:1982cy,Abbott:1982af,Dine:1982ah}. It was later argued that axions could form a Bose-Einstein condensate and exhibit superfluidity \cite{Sin:1992bg,Silverman:2002qx,Hu:2000ke,Sikivie:2009qn,Hui:2016ltb,Berezhiani:2015bqa,Ferreira:2018wup}.

To emphasize the superfluid nature of axions, we can rewrite the field equations in terms of fluid equations. Being comprised of extremely light particles at incredibly high number density, axion dark matter is well described by a coherent scalar field. Moreover, because dark matter as we observe it is \emph{cold}, the system is well described by a non-relativistic limit. The Euler and continuity equations of classical fluid mechanics emerge in this non-relativistic limit, defined via the decomposition
\be
\label{nrlim}
\varphi(x,t) = \sqrt{\frac{\hbar^3 c}{2m}} \left( \phi(x,t) e^{-i m c^2 t/\hbar} + c.c.\right) 
\ee
and the limit $|\ddot{\phi}| \ll m c^2 |\dot{\phi}|/\hbar$ \cite{Hui:2016ltb}. If we now define the fluid density $\rho$ and velocity ${\bf v}$ by:
\begin{equation}
\phi \equiv \sqrt{\frac{\rho}{m}} e^{i\theta} \quad , \quad {\bf
  v}\equiv \frac{\hbar}{R\,m} \nabla \theta
=\frac{\hbar}{2mi R}\left(\frac{1}{\phi}\nabla\phi -\frac{1}{\phi^*}\nabla\phi^*\right)\, 
\end{equation}
then the non-relativistic limit of the Klein-Gordon equation becomes,
\begin{align}
\label{massconserv}
\dot\rho + 3 H \rho + \frac{1}{R}\nabla \cdot (\rho {\bf v}) &= 0 \, , \\
\label{Euler} 
\dot {{\bf v}} + H {\bf v} + \frac{1}{R} ({\bf v} \cdot \nabla) {\bf v} &= -\frac{1}{R} \nabla \Phi + \frac{\hbar^2}{2 R^3 m^2} \nabla
\left( \frac{\nabla^2 \sqrt{\rho}}{\sqrt{\rho}} \right) , \nonumber \, 
\end{align}
These are the Madelung equations in an expanding universe, which are the continuity and Euler equations of fluid mechanics, with the addition of the second term on the right of the lower equation, referred to as the quantum pressure.

Parallel to the development of axions has been the study of condensate phases of non-Abelian gauge theories such as the Standard Model's Quantum Chromo-Dynamics (QCD). Starting from the realization that neutrons can undergo a BCS transition to a superfluid in the interior of neutron stars \cite{1969Natur.224673B}, it was found that at high enough densities, the quarks themselves could form Cooper pairs and undergo a BCS transition to a superfluid or superconducting state \cite{Alford:1997zt} (for a review see e.g.\cite{Alford:2007xm}). These developments have spurred on the study of neutron star physics (for reviews see \cite{Lombardo:2000ec,Dean:2002zx,Page:2013hxa,Haskell:2017lkl}), which with the observation of gravitational waves from a neutron star binary merger \cite{TheLIGOScientific:2017qsa}, may be on the cusp of a breakthrough.

Bringing together these disparate developments, it was shown in \cite{Alexander:2018fjp} that a QCD-like theory could lead to superfluidity on cosmological scales, and constitute a scenario for superfluid dark matter, providing a BCS analog to the axion's BEC. The natural embedding of this scenario in inflationary cosmology, with the fundamental degrees of freedom produced in huge numbers as a side-effect of baryogenesis, leads to a scenario of dark matter with observables from all stages of the evolution of the universe.

What these variant scenarios of superfluid dark matter (SFDM) have in common is the existence of vortices. Let us now see explicitly how vortex structures come out of superfluid halos. As per Equation \eqref{nrlim}, we can model superfluid dark matter with a macroscopic complex scalar function $\phi(r)$ that is described by the Lagrangian,
\begin{equation}
    L = \int {\rm d}^3 x \left( \frac{\hbar^2}{2m}\left| \nabla\phi \right|^2 + \frac{1}{4}\lambda \left|\phi\right|^4 + \frac{1}{2}m\Phi \left| \phi\right|^2 \right) ,
    \label{eq:1p1}
\end{equation}
The first term on the r.h.s of Equation \ref{eq:1p1} is the kinetic term, the second an effective interaction potential with coupling strength $\lambda = 4\pi \hbar^2 a/m$ where $a$ is the s--wave scattering length with a cross section $\sigma = 8\pi a^2$, and the last term is the coupling to the Newtonian gravitational potential $\Phi$. This system is completely described by the time--dependent Gross-Pitaevski equation and the Poisson equation,

\begin{eqnarray}
 i\hbar\dot{\phi} &&= \left(-\dfrac{\hbar^2}{2m}\nabla^{2}+m \Phi- \dfrac{\lambda}{m^{2}} \vert\phi\vert^{2}\right)\phi , \\
\nabla^{2}\Phi &&=4 \pi G m \vert\phi\vert^{2} ,
\end{eqnarray} 
where $\nabla^2$ is the spatial Laplacian and we take $c=1$.
The wave function $\phi$ can be written in terms of the modulus and phase,
\begin{equation}
    \phi(r,t) = \left| \phi \right|(r,t) e^{iS(r,t)},
\end{equation}
in terms of which the equation of motion becomes two equations,
\begin{eqnarray}
   -\frac{2m}{\hbar}\left| \phi \right|\frac{\partial S}{\partial t} + \Delta \left| \phi \right| -  \left| \phi \right| \left( \nabla S\right)^2 &&= \frac{2m}{\hbar^2}\left( m\Phi + g\left| \phi \right|^2\right)\left| \phi \right|  , \nonumber \\
     \frac{\partial \left| \phi \right|^2}{\partial t} + \nabla \cdot \left[ \left| \phi \right|^2 \frac{\hbar}{m} \nabla S\right] && = 0.
\end{eqnarray}
These equations form a system of quantum--mechanical hydrodynamic equations with a bulk velocity $v = \hbar/m \nabla S$. 

Note that the curl of the bulk velocity is null. This seems to imply that a SFDM halo would not have angular momentum. However, this would not be the case in reality. Thus the bulk velocity must contain a singularity; this  is the vortex. Note that an integral over a closed contour around such a vortex would be non--zero
\begin{equation}
    \oint_C \nabla S \cdot dl = 2\pi d \frac{\hbar}{m} .
\end{equation}
Here $d$ is an integer called the winding number. This implies that angular momentum of the vortex is quantized. 

Solutions describing vortices in dark matter halos were found in \cite{sfdm}. The vortex solution is characterized by a density profile that can be parameterized as \cite{sfdm},
\[
    \rho_{v}(r,z)= 
\begin{cases}
    0 ,&  r > r_{v}\\
    \rho_{v0}\left[ \left(\frac{r}{r_{v}}\right)^{\alpha_v}-1 \right],              &  r \leq r_v
\end{cases}
\]
where $r$ is the radial coordinate in cylindrical coordinates, $r_v$ is the core-radius of the vortex solution, and $\alpha_v$ is a scaling exponent. This effectively models the vortex as a tube. On distance scales much larger then $r_v$, the vortex can be approximated as a line of density $\rho_{v0}$.

\begin{figure*}
\centering
\begin{minipage}{.45\linewidth}
  \includegraphics[width=.85\linewidth]{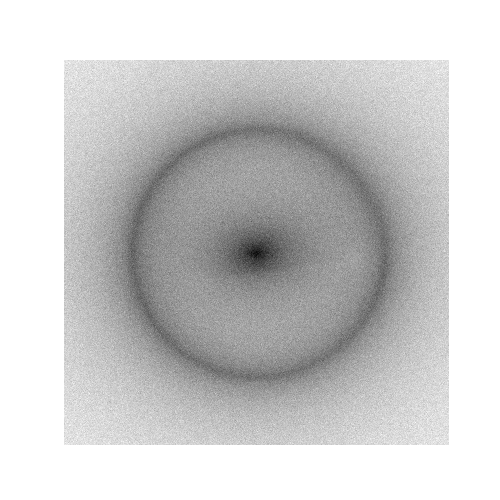}%
  
\caption{\label{fig:lens} Lens image with superfluid substructure (a vortex). Simulated with the \textit{PyAutoLens} software suite.}
\end{minipage}
\hspace{.05\linewidth}
\begin{minipage}{.45\linewidth}
  \includegraphics[width=.85\linewidth]{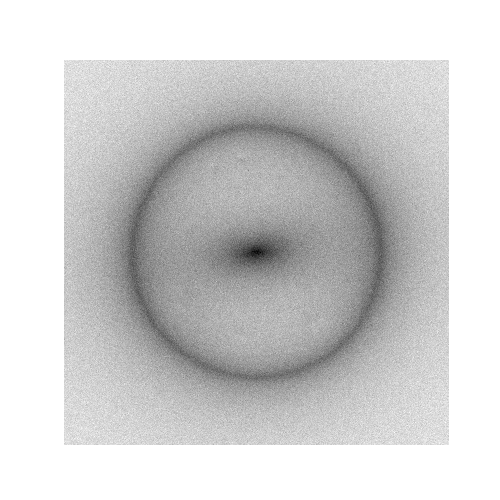}
  \caption{Same as Figure \ref{fig:lens} but for particle substructure.}
  \label{fig:lens2}
\end{minipage}
\end{figure*}

\begin{figure*}
\centering
\begin{minipage}{.45\linewidth}
  \includegraphics[width=1.\linewidth]{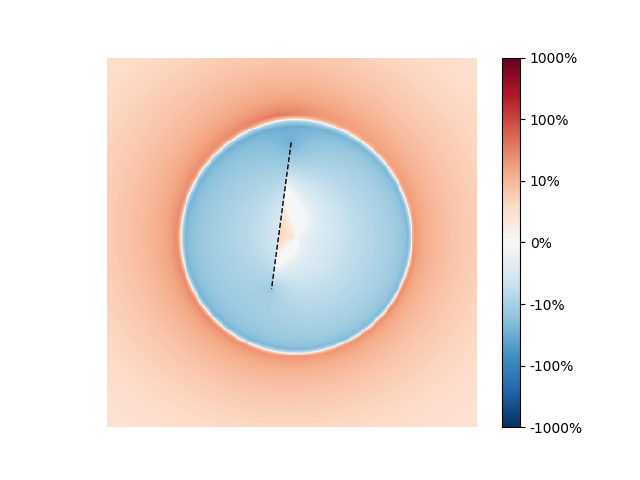}
  \caption{Residuals image for superfluid-like substructure. Here 1\% of the halo mass. Dashed line represents position of the vortex.}
  \label{img1}
\end{minipage}
\hspace{.05\linewidth}
\begin{minipage}{.45\linewidth}
  \includegraphics[width=1.\linewidth]{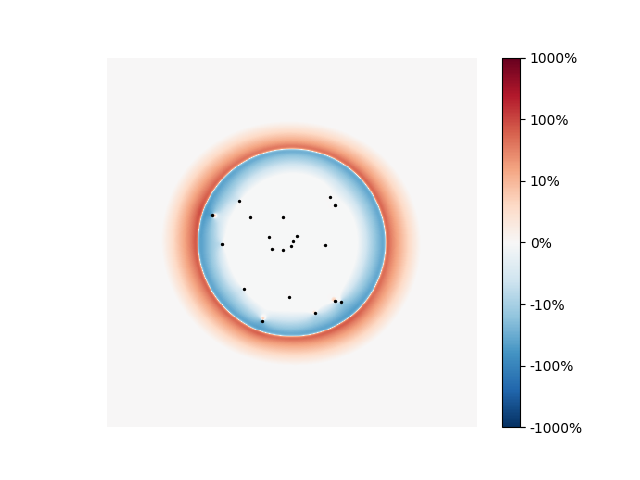}
  \caption{Residuals image for particle dark matter-like substructure. Here 1\% of the halo mass. Black dots represent positions of subhalos.}
  \label{img2}
\end{minipage}
\end{figure*}

The values of these parameters, most importantly the density and total mass of the vortices, as well as the expected number density in realistic dark matter halos, varies widely across the literature. For example, the total amount of vortices in halos range from 340 vortices in the M31 halo with assumed constituent particle mass $m = 10^{-23}$ eV \cite{Silverman:2002qx} to $N = 10^{23}$ vortices in a typical DM halo for $m = 1$ eV \cite{Berezhiani:2015bqa}. In \cite{Banik:2013rxa} it is shown that vortices can have mutual attraction and, over time, coalesce into a single, more massive vortex. 

We also note that other substructure exists in superfluid scenarios, such as the recently found thin-disk solutions \cite{Alexander:2019qsh}. These have a lensing signal which interpolates between that of a vortex and a spherical halo, depending on the orientation of the disk. 

Finally, we note the relation of vortices to \emph{cosmic strings}. The latter is often explained as the relativistic analog to vortices (see e.g. \cite{Brandenberger:1993by,Brandenberger:2013tr}), formed during a phase transition in a relativistic quantum field theory. As such, cosmic stings have a transverse velocity that is close to the speed of light. In spite of this, much of the work on strong lensing by cosmic strings \cite{Sazhin:2006kf,Gasparini:2007jj,Morganson:2009yk} has approximated them as stationary or non-relativistic, and hence effectively behaving as vortices. While we do not use these results directly, the lensing images generated in the following section agree with the results obtained in the cosmic string literature.

\section{Strong Lensing Images}
\label{sec:stronglensing}

\begin{figure*}
    \centering
    \includegraphics[scale=0.4, trim = {0 4.6cm 0 4.5cm}]{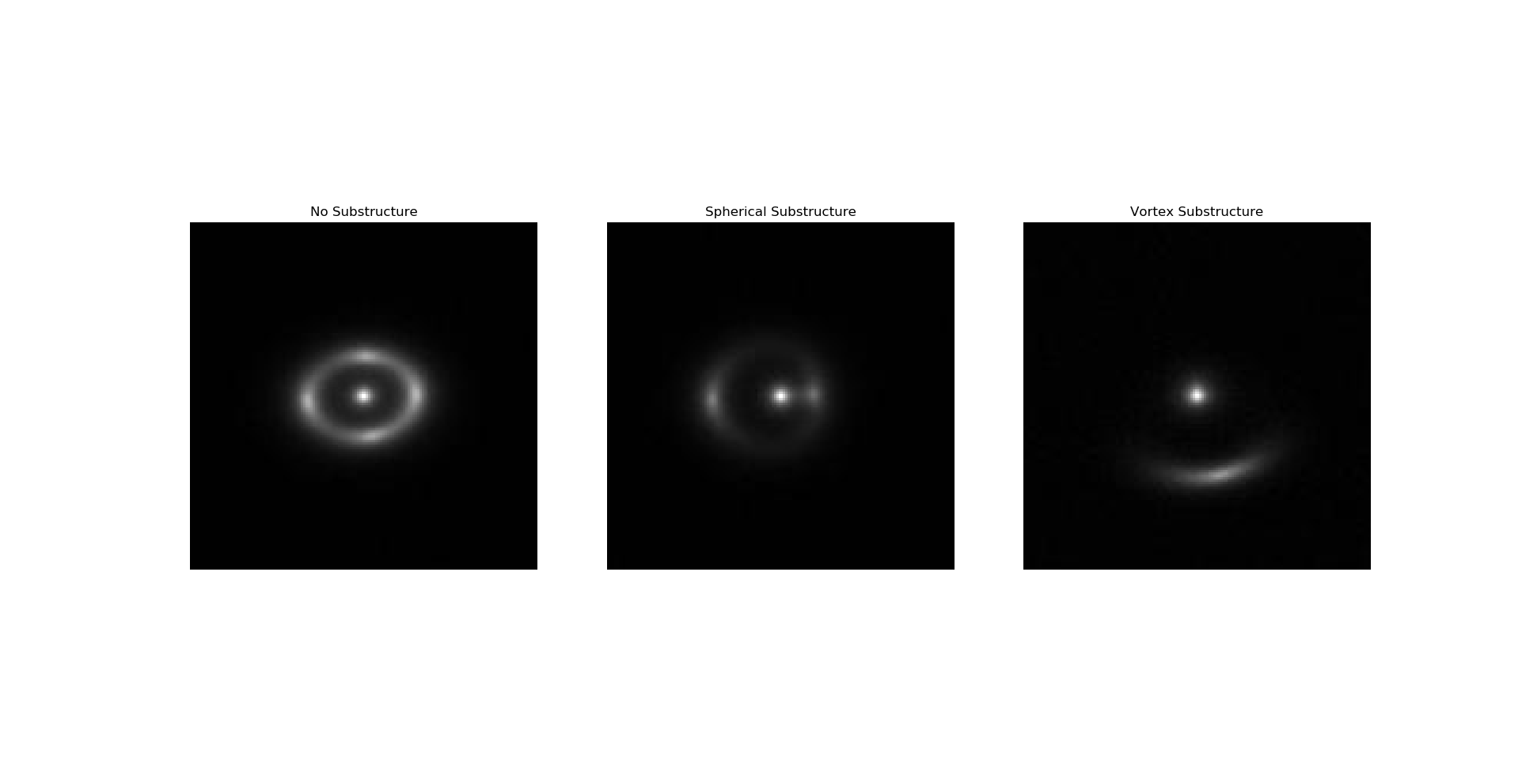}
    \caption[Model C: Simulated Images]{Example simulated images for all three classes.}
    \label{fig:modC}
\end{figure*}

At this moment strong lensing data is limited to a handful of images. However, the upcoming completion of the Large Synoptic Survey Telescope (LSST) will lead to thousands of strong lensing images that can be analyzed \cite{lsstw}. In this work we have chosen to simulate our lensing images using the package \textit{PyAutoLens} \cite{night, night2}. Written in Python, it can produce a variety of simulated strong lensing images where the user can adjust, among many possibilities, the mass of the halo, include substructure, light profiles, and mass profiles.

In addition to the simulation of the lensing itself, we also consider the addition of noise and the modifications induced by a point spread function (PSF) on our observation. Thus, we can vary the level of noise in our images and include a PSF that is inline with real world instruments like Hubble or the future LSST, in this case both sub--arcsecond resolution. Following  \cite{pcat}, we approximate the PSF as an Airy disk
whose first zero-crossing occurs at a radius of $\sigma_{psf} \lesssim {\rm arcsec}$. This approximation is valid when with noise is dominated by diffraction, which we assume to be the case.

The lensing image due to a single vortex embedded in a halo, with the vortex mass 1\% that of the halo, is shown in Figure \ref{fig:lens}. We do the same for spherical substructure, as studied in \cite{pcat}, in Figure \ref{fig:lens2}. To quantify the effect of the substructure we subtract from each image the lensing image due the halo alone, and show the result (the `residuals') in Figure \ref{img1} and Figure \ref{img2} for vortex and spherical substructure respectively. From these images one can appreciate the difference in lensing is primarily in the the \emph{morphology} of the signal, making this an ideal task for classification with a convolutional neural network.

Lensing images were generated with parameters and their distributions given in Table \ref{tab:table}.  We have included the light from the lensing galaxy and non--negligible backgrounds and noise. We have also accounted for other instrumental effects like the point--spread--function which we have modeled after the expected resolution of LSST, as well as shear effects. Example images can be found for each class in Figure \ref{fig:modC}.

\section{Network \& Training}
\label{CNN}

In this work we take a supervised approach to establish a set of performance benchmarks for identifying different types of dark matter substructure. We, therefore, simulate the expected lensing effects from a variety of substructures and train a CNN to distinguish among them.

As the total mass constrained in the substructure is likely a small fraction of the total lensing mass, it may prove challenging to distinguish dark matter with traditional methods.

The addition of noise and other astrophysical backgrounds makes this an even more challenging task. Deep learning methods are more amenable to identifying subtle morphologies in images. This is the approach we take in this work.

Convolutional neural networks (CNN) are the natural choice for working with images. There are several pre-trained networks openly available, e.g. \textit{ResNet} (or \textit{residual neural network})\cite{DBLP:journals/corr/HeZRS15}, \textit{AlexNet} \cite{Krizhevsky:2017:ICD:3098997.3065386}, \textit{DenseNet} \cite{DBLP:journals/corr/HuangLW16a}, and \textit{VGG} \cite{2014arXiv1409.1556S}. For \textit{ResNet}, the defining feature is that residual networks can skip layers all together in training. This, in practice, helps speed up the learning rate of the network by allowing the network to train fewer layers in the initial stages of learning. For this reason we will focus on \textit{ResNet}, and return to a detailed algorithm comparison in section VI.


During training we make use of data augmentation (see e.g. \cite{Krizhevsky:2017:ICD:3098997.3065386}) via translation and rotations up to 90$^\circ$. These all constitute invariant transformations with respect to the underlying substructure that allow the network to learn the actual structure in images. Thus data augmentation aims to increase our learning efficiency by seeing the same image in new ways several times.

We utilize 150,000 training and 15,000 validation images. The binary cross--entropy loss was minimized with the Atom optimizer in batches of 200 over a total of at most 20 epochs. The learning rate starts with a value of  $1 \times 10^{-4}$ and is reduced by a factor of ten when the validation loss is not improved for 3 consecutive epochs. The networks were implemented using the PyTorch package and run on a single NVIDIA Titan K80 GPU. We use the well-established area under the ROC curve (AUC) as a metric for classifier performance

\section{Results}
\label{results}

As discussed in previous sections, we are interested in identifying and classifying substructure in strong gravitational lensing images. We do this with a supervised CNN, which requires that the classes be identified from the outset. In addition to the vortex and spherical sub-halo substructure classes discussed in Section \ref{sec:stronglensing}, there remains the possibility that an image may not have any detectable substructure at all, e.g. if the Einstein radius of the substructure is predominantly smaller then the PSF of the detector. Given this, we introduce an additional class: no substructure present.

In what follows we will train a multi--class classifier to predict the three classes: vortex, spherical, and no-substructure. We train three additional binary classifiers to distinguish between the two most probable classes predicted by the multi-class classifier. We do this using realistic mock lensing images as described in section \ref{sec:stronglensing}, with parameters and their distributions given in Table \ref{tab:table}.

We start by considering an idealized population of physical systems, with e.g. the distance to the lensed and lensing galaxy the same in each image. This could plausibly be the case if a very large data set was first divided into subsets exhibiting roughly constant properties. This is useful for comparison to \cite{pcat}, which held fixed the number of sub-halos, and as a playground for performance tests of differing network architectures.  We then consider a less-idealized population of images, with several additional physical properties of the halo and substructure allowed to vary. The parameters for each case are given in Table \ref{tab:table}.

\subsection{A Multi-Class Classifier for Substructure Morphology}
\label{sec:modela}

To begin, we consider similar parameters as images simulated in \cite{pcat}, which used probabilistic cataloging to identify spherical substructure. We differ with \cite{pcat} by modeling the spherical substructure as point masses as opposed to truncated--NFW profiles, and of course by the inclusion of vortex--like substructure. Similar to \cite{pcat}, we have chosen a fixed value of 25 subhalos.

\begin{figure}
    \centering
    \includegraphics[width=0.45 \textwidth]{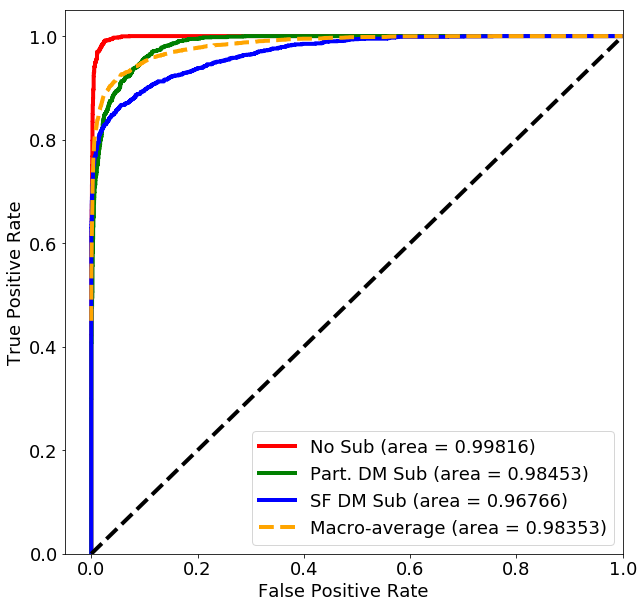}
    \caption{ROC curve for multiclass substructure classification with \textit{ResNet}, as discussed in Section \ref{sec:modela}.}
  \label{imga}
\end{figure}

We use a pre--trained network as described in Section \ref{CNN}. The ROC-AUC curves for substructure classification by \textit{ResNet} are shown in Figure \ref{imga}. As can be appreciated from the AUC scores of 0.998, 0.985, and 0.968, for images with no substructure, spherical sub-halos, and vortices, respectively, our algorithm achieves excellent classification.

\begin{figure}
  \includegraphics[width=0.5 \textwidth]{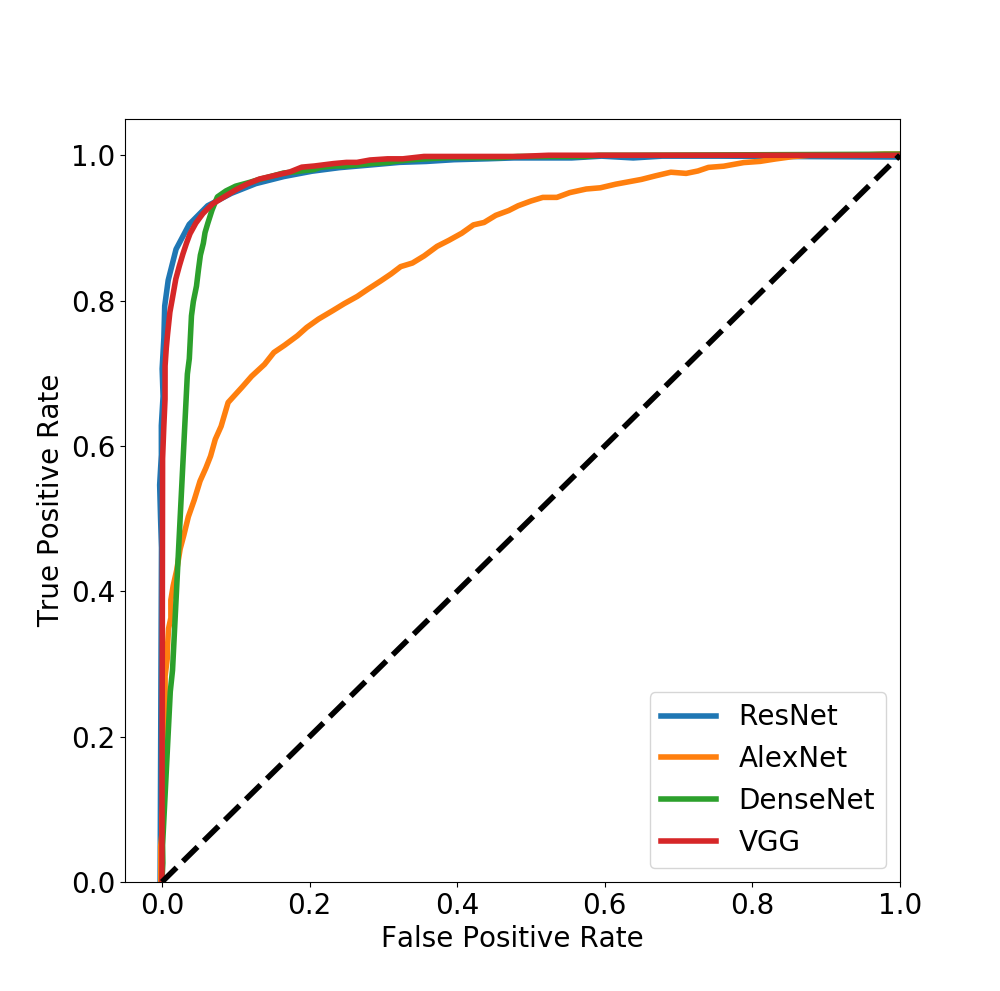}
  \caption[ROC -- Multiclass Classifier]{Comparison of architechures. Shown is the ROC curve averaged across substructure types for \textit{Resnet}, \textit{Alexnet}, \textit{VGG}, and \textit{DenseNet}.}
  \label{fig:compare}
\end{figure}

For the sake of comparison, we repeat this with differing choices of architecture, namely \textit{VGG}, \textit{DenseNet}, and \textit{AlexNet}. The resulting ROC curves (averaged over substructure types) are shown in Figure \ref{fig:compare}. While \textit{ResNet}, \textit{VGG} and \textit{DenseNet} show comparable performance, \textit{AlexNet} performs considerably worse. The other three architectures were otherwise indistinguishable, with the exception that, as expected,  \textit{ResNet} is more computationally efficient than both \textit{VGG} and \textit{DenseNet}.

\begin{table*}
    \scriptsize
	\centering
	\caption[Simulation Parameters]{Parameters with distributions and priors used in the simulation of strong lensing images. Where two values are given, the first corresponds to section VI.A. and the second corresponds to VI.B. Note that only a single type of substructure was used per image.}
	\label{tab:table}
	\begin{tabular}{cccc} 
		\hline
		\hline
		Lensing Galaxy -- \textit{Sersic Light Profile}\\
		\hline
		\hline
		\textbf{Parameter} & \textbf{Distribution} & \textbf{Priors} & \textbf{Details} \\
		\hline
$\theta_x$  &  fixed & 0 & x position \\
$\theta_y$  &  fixed & 0 & y position\\
z  &  fixed $\mid$ uniform & 0.5 $\mid$ [0.4,0.6]  & redshift\\
e  &  uniform & [0.5, 1.0] & axis ratio\\
$\phi$ & uniform & [0, 2$\pi$] & orientation relative to y axis \\
I & fixed & 1.2  & intensity of emission (arbitrary units) \\
n & fixed & 2.5 & Sersic index \\
R & fixed $\mid$ uniform & 0.5 $\mid$ [0.5,2] & effective radius \\
		\hline
		\hline
		Dark Matter Halo -- \textit{Spherical Isothermal}\\
		\hline
		\hline
		\textbf{Parameter} & \textbf{Distribution} & \textbf{Priors} & \textbf{Details} \\
		\hline
$\theta_x$  &  fixed & 0 & x position \\
$\theta_y$  &  fixed & 0 & y position \\
$\theta_E$ & fixed & 1.2 & Einstein radius\\
        \hline
		\hline
		External Shear\\
		\hline
		\hline
		\textbf{Parameter} & \textbf{Distribution} & \textbf{Priors} & \textbf{Details} \\
		\hline
		$\gamma_{ext}$ & uniform & [0.0, 0.3] & magnitude \\
		$\phi_{ext}$ & uniform & [0, 2$\pi$] & angle \\
        \hline
		\hline
		Lensed Galaxy -- \textit{Sersic Profile}\\
		\hline
		\hline
		\textbf{Parameter} & \textbf{Distribution} & \textbf{Priors} & \textbf{Details} \\
		\hline
$r$ & uniform & [0, 1.2] & radial distance from center\\
$\phi_{bk}$ & uniform & [0, 2$\pi$] & angular position of galaxy from y axis\\ 
z  &  fixed $\mid$ uniform & 1.0 $\mid$ [0.8,1.2]  & redshift\\
e  &  uniform & [0.7, 1.0] & axis ratio\\
$\phi$ & uniform & [0, 2$\pi$] & orientation relative to y axis \\
I & uniform & [0.7, 0.9] & intensity of emission (arbitrary units) \\
n & fixed & 1.5 & Sersic index \\
R & fixed & 0.5 & effective radius \\
        \hline
		\hline
		Vortex \\
		\hline
		\hline
		\textbf{Parameter} & \textbf{Distribution} & \textbf{Priors} & \textbf{Details} \\
		\hline
$\theta_x$  &  fixed $\mid$ normal & 0 $\mid$ $[0.0,0.5]$ & x position \\
$\theta_y$  &  fixed $\mid$ normal & 0 $\mid$ $[0.0,0.5]$ & y position\\
$l$ & fixed $\mid$ uniform & 1.0 $\mid$ [0.5,2.0] & length of vortex\\
$\phi_{v}$ & uniform & [0, 2$\pi$] & orientation from y axis\\ 
$m_{vort}$ & fixed & 0.01 $M_{Halo}$ & total mass of vortex\\
        \hline
		\hline
		Spherical \\
		\hline
		\hline
		\textbf{Parameter} & \textbf{Distribution} & \textbf{Priors} & \textbf{Details} \\
		\hline
$r$ & uniform & [0, 1.0] & radial distance from center\\
$\phi_{sph}$ & uniform & [0, 2$\pi$] & angular position of galaxy from y axis\\ 
$N$ & fixed $\mid$ Poisson & 25  $\mid$ $\mu$=25 & number of substructures\\
$m_{sub}$ & fixed & 0.01 $M_{Halo}$ & total mass of subhalos\\

	\end{tabular}
\end{table*}

\subsection{Towards a Representative Population of Images}
\label{sec:modelb}

We now allow the training data to be a more diverse set of physical systems, as may be the case with actual data. We allow for variation in the distance to the lensing and lensed galaxy,  the galaxy size, and importantly, the intensity of the background and noise, allowing the background to become non-negligible.  With regards to substructure, we vary the position of the vortex, and for spherical substructure consider the number of halos to be taken from a Poisson draw with mean 25. The details of all parameters and the specific distributions from which we draw values for simulation are all included in Table \ref{tab:table}.

\begin{figure}
  \centering
  \includegraphics[width=0.45 \textwidth]{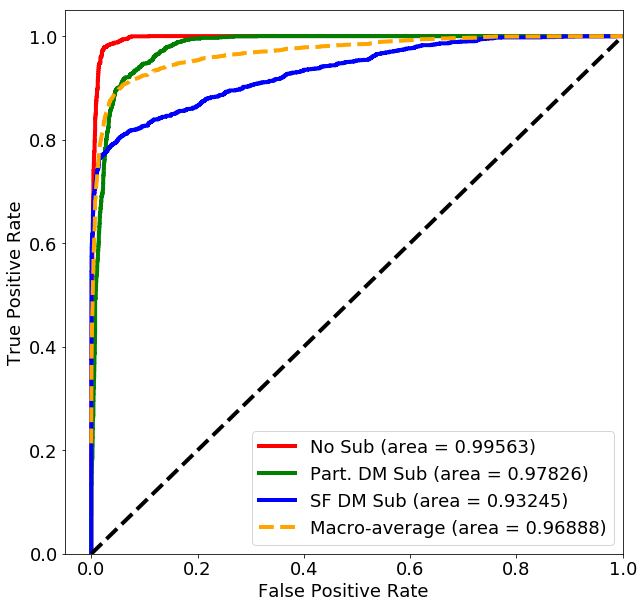}
  \caption{ROC curve for multiclass substructure classification with \textit{ResNet}, including additional variations across the population of images, as discussed in Section \ref{sec:modelb}.}
  \label{imgb}
\end{figure}

Utilizing the same \emph{ResNet} architecture and running for 20 epochs while updating the learning rates, training our multiclass classifier obtains good results with a macro--averaged AUC of $0.969$. The ROC curve is shown in Figure \ref{imgb}. 

\section{Towards the Detection of Substructure}
\label{detection}

To complete the analysis of this work, we establish the detection threshold for our network. To do so, we change the total mass of the substructure while holding all other parameters constant. Of course, in practice, it would be possible to train for more epochs, use a deeper network, add more training images, etc., and push our threshold further. We implement this by simulating sets of 50,000 training and 5,000 validation images at different total fractions of the halo mass for each class. We train each set on the same architecture, here \textit{ResNet}, for 10 epochs. The metric we use to parameterize the ability of the network to learn is the AUC.

\begin{figure}[b]
\includegraphics[width= 0.45\textwidth]{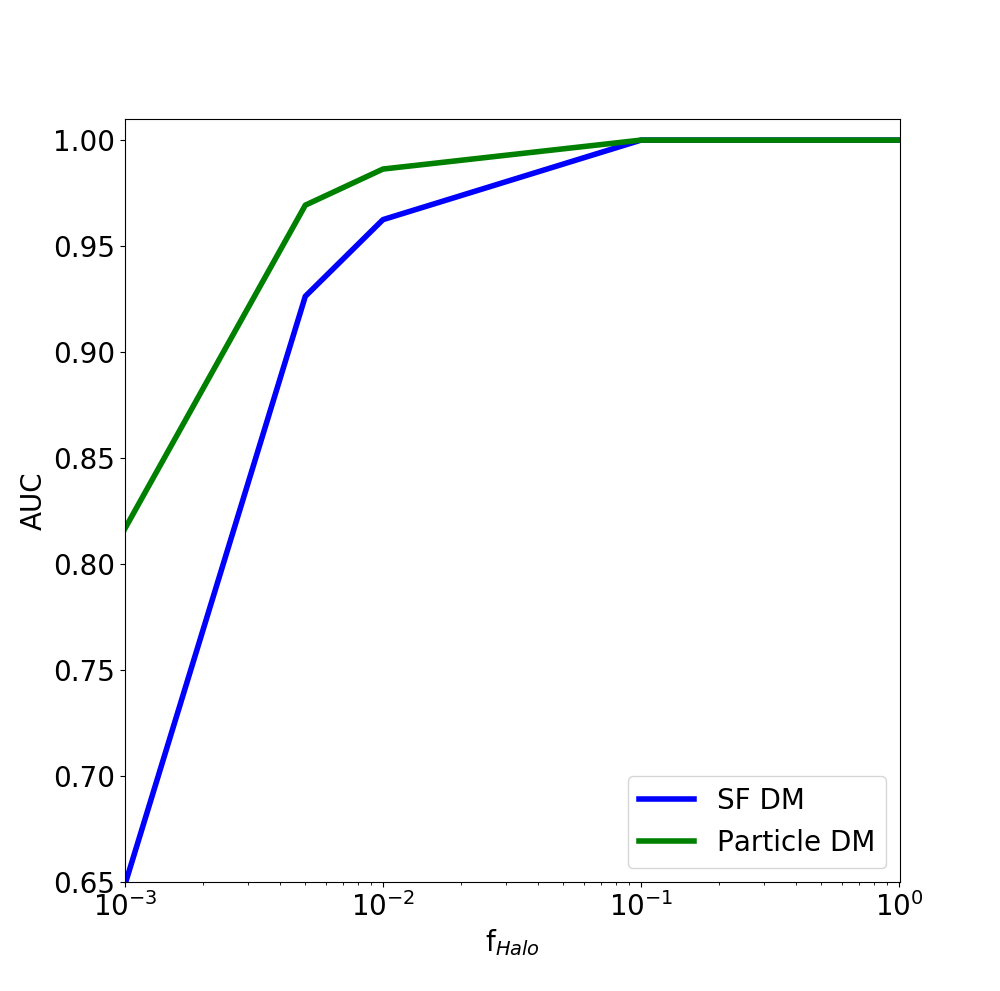}
\caption{\label{fig:thresh} AUC as a function of the ratio of substructure to host halo mass.}
\end{figure}

The AUC score for varying fraction of the halo mass contained in both types of substructure is shown in Figure \ref{fig:thresh}. From this one can appreciate that the AUC rapidly deteriorates for a substructure mass below $10^{-2.5} \approx 0.3\%$ of the halo mass. From this we conclude that a CNN, given the fixed computing resources stated above, can reliably identify lensing images containing a substructure provided that its collective mass constitutes at least a fraction of a percent of the dark matter in the halo.

\vspace{1cm} 

\section{Discussion \& Conclusion}
\label{discussion}

It is well established that substructure can constrain dark matter models. In this work we have proposed it may even identify the nature of dark matter. Motivated by the significant theoretical uncertainty as to the nature of dark matter, in this work we have considered the study of substructure as a \emph{classification problem}, and investigated the feasibility of using a machine learning architecture to distinguish different types of substructure in strong lensing images.

Utilizing a simple supervised convolutional neural network, trained on simulated images, we have demonstrated that it is indeed feasible for a network to reliably distinguish among different types of dark matter substructure.   This compliments existing approaches to substructure, namely the statistical detection \cite{Rivero:2017mao,Cyr-Racine:2015jwa,Cyr-Racine:2018htu,Rivero:2018bcd,Brennan:2018jhq} and the pinpointing of individual substructures (e.g. \cite{alma}), and could be used as a part of a data analysis pipeline in the latter task.

The success of the supervised approach utilized here provides confidence in future implementations of an \emph{unsupervised architecture}. This would allow the analysis to be fully agnostic as to the true nature of dark matter, in recognition of the possibility that dark matter could be outside the scope of current theoretical expectations.  

One promising path in this direction is to implement an auto-encoder to learn the underlying substructure of real images. With thousands of galaxy--galaxy strong lensing images expected in the next few years \cite{lsstw}, there should be ample data for training. Given this, and based on the results in this work, an unsupervised approach certainly looks very promising.

Finally, we note that deep learning may be amenable to searching for dark matter vortices in other observational windows, analogous to searches for cosmic strings in the cosmic microwave background \cite{Ciuca:2017gca,Ciuca:2018tei} and 21cm \cite{Brandenberger:2010hn,McDonough:2011er}. We leave this, and the development of an unsupervised approach, to future work.

\begin{acknowledgments}
The authors thank Cora Dvorkin, Javad Hashemi, Shirley Ho, and David Spergel, for useful discussions. One of the authors thanks Robert Brandenberger for encouragement to work on this topic more than 20 years ago.

\end{acknowledgments}

\bibliography{apssamp}

\begin{thebibliography}{84}%
\makeatletter
\providecommand \@ifxundefined [1]{%
 \@ifx{#1\undefined}
}%
\providecommand \@ifnum [1]{%
 \ifnum #1\expandafter \@firstoftwo
 \else \expandafter \@secondoftwo
 \fi
}%
\providecommand \@ifx [1]{%
 \ifx #1\expandafter \@firstoftwo
 \else \expandafter \@secondoftwo
 \fi
}%
\providecommand \natexlab [1]{#1}%
\providecommand \enquote  [1]{``#1''}%
\providecommand \bibnamefont  [1]{#1}%
\providecommand \bibfnamefont [1]{#1}%
\providecommand \citenamefont [1]{#1}%
\providecommand \href@noop [0]{\@secondoftwo}%
\providecommand \href [0]{\begingroup \@sanitize@url \@href}%
\providecommand \@href[1]{\@@startlink{#1}\@@href}%
\providecommand \@@href[1]{\endgroup#1\@@endlink}%
\providecommand \@sanitize@url [0]{\catcode `\\12\catcode `\$12\catcode
  `\&12\catcode `\#12\catcode `\^12\catcode `\_12\catcode `\%12\relax}%
\providecommand \@@startlink[1]{}%
\providecommand \@@endlink[0]{}%
\providecommand \url  [0]{\begingroup\@sanitize@url \@url }%
\providecommand \@url [1]{\endgroup\@href {#1}{\urlprefix }}%
\providecommand \urlprefix  [0]{URL }%
\providecommand \Eprint [0]{\href }%
\providecommand \doibase [0]{http://dx.doi.org/}%
\providecommand \selectlanguage [0]{\@gobble}%
\providecommand \bibinfo  [0]{\@secondoftwo}%
\providecommand \bibfield  [0]{\@secondoftwo}%
\providecommand \translation [1]{[#1]}%
\providecommand \BibitemOpen [0]{}%
\providecommand \bibitemStop [0]{}%
\providecommand \bibitemNoStop [0]{.\EOS\space}%
\providecommand \EOS [0]{\spacefactor3000\relax}%
\providecommand \BibitemShut  [1]{\csname bibitem#1\endcsname}%
\let\auto@bib@innerbib\@empty
\bibitem [{\citenamefont {Drukier}\ \emph {et~al.}(1986)\citenamefont
  {Drukier}, \citenamefont {Freese},\ and\ \citenamefont
  {Spergel}}]{Drukier:1986tm}%
  \BibitemOpen
  \bibfield  {author} {\bibinfo {author} {\bibfnamefont {A.~K.}\ \bibnamefont
  {Drukier}}, \bibinfo {author} {\bibfnamefont {K.}~\bibnamefont {Freese}}, \
  and\ \bibinfo {author} {\bibfnamefont {D.~N.}\ \bibnamefont {Spergel}},\
  }\href {\doibase 10.1103/PhysRevD.33.3495} {\bibfield  {journal} {\bibinfo
  {journal} {Phys. Rev.}\ }\textbf {\bibinfo {volume} {D33}},\ \bibinfo {pages}
  {3495} (\bibinfo {year} {1986})}\BibitemShut {NoStop}%
\bibitem [{\citenamefont {Goodman}\ and\ \citenamefont
  {Witten}(1985)}]{Goodman:1984dc}%
  \BibitemOpen
  \bibfield  {author} {\bibinfo {author} {\bibfnamefont {M.~W.}\ \bibnamefont
  {Goodman}}\ and\ \bibinfo {author} {\bibfnamefont {E.}~\bibnamefont
  {Witten}},\ }\href {\doibase 10.1103/PhysRevD.31.3059} {\bibfield  {journal}
  {\bibinfo  {journal} {Phys. Rev.}\ }\textbf {\bibinfo {volume} {D31}},\
  \bibinfo {pages} {3059} (\bibinfo {year} {1985})},\ \bibinfo {note}
  {[,325(1984)]}\BibitemShut {NoStop}%
\bibitem [{\citenamefont {Akerib}\ \emph {et~al.}(2017)\citenamefont {Akerib}
  \emph {et~al.}}]{Akerib:2016vxi}%
  \BibitemOpen
  \bibfield  {author} {\bibinfo {author} {\bibfnamefont {D.~S.}\ \bibnamefont
  {Akerib}} \emph {et~al.} (\bibinfo {collaboration} {LUX}),\ }\href {\doibase
  10.1103/PhysRevLett.118.021303} {\bibfield  {journal} {\bibinfo  {journal}
  {Phys. Rev. Lett.}\ }\textbf {\bibinfo {volume} {118}},\ \bibinfo {pages}
  {021303} (\bibinfo {year} {2017})},\ \Eprint
  {http://arxiv.org/abs/1608.07648} {arXiv:1608.07648 [astro-ph.CO]}
  \BibitemShut {NoStop}%
\bibitem [{\citenamefont {Cui}\ \emph {et~al.}(2017)\citenamefont {Cui} \emph
  {et~al.}}]{Cui:2017nnn}%
  \BibitemOpen
  \bibfield  {author} {\bibinfo {author} {\bibfnamefont {X.}~\bibnamefont
  {Cui}} \emph {et~al.} (\bibinfo {collaboration} {PandaX-II}),\ }\href
  {\doibase 10.1103/PhysRevLett.119.181302} {\bibfield  {journal} {\bibinfo
  {journal} {Phys. Rev. Lett.}\ }\textbf {\bibinfo {volume} {119}},\ \bibinfo
  {pages} {181302} (\bibinfo {year} {2017})},\ \Eprint
  {http://arxiv.org/abs/1708.06917} {arXiv:1708.06917 [astro-ph.CO]}
  \BibitemShut {NoStop}%
\bibitem [{\citenamefont {Aprile}\ \emph {et~al.}(2018)\citenamefont {Aprile}
  \emph {et~al.}}]{Aprile:2018dbl}%
  \BibitemOpen
  \bibfield  {author} {\bibinfo {author} {\bibfnamefont {E.}~\bibnamefont
  {Aprile}} \emph {et~al.} (\bibinfo {collaboration} {XENON}),\ }\href
  {\doibase 10.1103/PhysRevLett.121.111302} {\bibfield  {journal} {\bibinfo
  {journal} {Phys. Rev. Lett.}\ }\textbf {\bibinfo {volume} {121}},\ \bibinfo
  {pages} {111302} (\bibinfo {year} {2018})},\ \Eprint
  {http://arxiv.org/abs/1805.12562} {arXiv:1805.12562 [astro-ph.CO]}
  \BibitemShut {NoStop}%
\bibitem [{\citenamefont {Aaboud}\ \emph {et~al.}(2019)\citenamefont {Aaboud}
  \emph {et~al.}}]{Aaboud:2019yqu}%
  \BibitemOpen
  \bibfield  {author} {\bibinfo {author} {\bibfnamefont {M.}~\bibnamefont
  {Aaboud}} \emph {et~al.} (\bibinfo {collaboration} {ATLAS}),\ }\href
  {\doibase 10.1007/JHEP05(2019)142} {\bibfield  {journal} {\bibinfo  {journal}
  {JHEP}\ }\textbf {\bibinfo {volume} {05}},\ \bibinfo {pages} {142} (\bibinfo
  {year} {2019})},\ \Eprint {http://arxiv.org/abs/1903.01400} {arXiv:1903.01400
  [hep-ex]} \BibitemShut {NoStop}%
\bibitem [{\citenamefont {{A. Burkert}}(1996)}]{burk}%
  \BibitemOpen
  \bibfield  {author} {\bibinfo {author} {\bibnamefont {{A. Burkert}}},\
  }\href@noop {} {\bibfield  {journal} {\bibinfo  {journal} {ApJ}\ }\textbf
  {\bibinfo {volume} {447}},\ \bibinfo {pages} {L25} (\bibinfo {year}
  {1996})},\ \bibinfo {note}
  {\href{https://arxiv.org/abs/astro-ph/9508025}{arXiv}}\BibitemShut {NoStop}%
\bibitem [{\citenamefont {{J.F. Navarro, C.S. Frenk and S.D.M.
  White}}(1996)}]{nfw}%
  \BibitemOpen
  \bibfield  {author} {\bibinfo {author} {\bibnamefont {{J.F. Navarro, C.S.
  Frenk and S.D.M. White}}},\ }\href@noop {} {\bibfield  {journal} {\bibinfo
  {journal} {ApJ}\ }\textbf {\bibinfo {volume} {462}},\ \bibinfo {pages} {563}
  (\bibinfo {year} {1996})},\ \bibinfo {note}
  {\href{https://arxiv.org/abs/astro-ph/9508025}{arXiv}}\BibitemShut {NoStop}%
\bibitem [{\citenamefont {Sin}(1994)}]{Sin:1992bg}%
  \BibitemOpen
  \bibfield  {author} {\bibinfo {author} {\bibfnamefont {S.-J.}\ \bibnamefont
  {Sin}},\ }\href {\doibase 10.1103/PhysRevD.50.3650} {\bibfield  {journal}
  {\bibinfo  {journal} {Phys. Rev.}\ }\textbf {\bibinfo {volume} {D50}},\
  \bibinfo {pages} {3650} (\bibinfo {year} {1994})},\ \Eprint
  {http://arxiv.org/abs/hep-ph/9205208} {arXiv:hep-ph/9205208 [hep-ph]}
  \BibitemShut {NoStop}%
\bibitem [{\citenamefont {Silverman}\ and\ \citenamefont
  {Mallett}(2002)}]{Silverman:2002qx}%
  \BibitemOpen
  \bibfield  {author} {\bibinfo {author} {\bibfnamefont {M.~P.}\ \bibnamefont
  {Silverman}}\ and\ \bibinfo {author} {\bibfnamefont {R.~L.}\ \bibnamefont
  {Mallett}},\ }\href {\doibase 10.1023/A:1015934027224} {\bibfield  {journal}
  {\bibinfo  {journal} {Gen. Rel. Grav.}\ }\textbf {\bibinfo {volume} {34}},\
  \bibinfo {pages} {633} (\bibinfo {year} {2002})}\BibitemShut {NoStop}%
\bibitem [{\citenamefont {Hu}\ \emph {et~al.}(2000)\citenamefont {Hu},
  \citenamefont {Barkana},\ and\ \citenamefont {Gruzinov}}]{Hu:2000ke}%
  \BibitemOpen
  \bibfield  {author} {\bibinfo {author} {\bibfnamefont {W.}~\bibnamefont
  {Hu}}, \bibinfo {author} {\bibfnamefont {R.}~\bibnamefont {Barkana}}, \ and\
  \bibinfo {author} {\bibfnamefont {A.}~\bibnamefont {Gruzinov}},\ }\href
  {\doibase 10.1103/PhysRevLett.85.1158} {\bibfield  {journal} {\bibinfo
  {journal} {Phys. Rev. Lett.}\ }\textbf {\bibinfo {volume} {85}},\ \bibinfo
  {pages} {1158} (\bibinfo {year} {2000})},\ \Eprint
  {http://arxiv.org/abs/astro-ph/0003365} {arXiv:astro-ph/0003365 [astro-ph]}
  \BibitemShut {NoStop}%
\bibitem [{\citenamefont {Sikivie}\ and\ \citenamefont
  {Yang}(2009)}]{Sikivie:2009qn}%
  \BibitemOpen
  \bibfield  {author} {\bibinfo {author} {\bibfnamefont {P.}~\bibnamefont
  {Sikivie}}\ and\ \bibinfo {author} {\bibfnamefont {Q.}~\bibnamefont {Yang}},\
  }\href {\doibase 10.1103/PhysRevLett.103.111301} {\bibfield  {journal}
  {\bibinfo  {journal} {Phys. Rev. Lett.}\ }\textbf {\bibinfo {volume} {103}},\
  \bibinfo {pages} {111301} (\bibinfo {year} {2009})},\ \Eprint
  {http://arxiv.org/abs/0901.1106} {arXiv:0901.1106 [hep-ph]} \BibitemShut
  {NoStop}%
\bibitem [{\citenamefont {Hui}\ \emph {et~al.}(2017)\citenamefont {Hui},
  \citenamefont {Ostriker}, \citenamefont {Tremaine},\ and\ \citenamefont
  {Witten}}]{Hui:2016ltb}%
  \BibitemOpen
  \bibfield  {author} {\bibinfo {author} {\bibfnamefont {L.}~\bibnamefont
  {Hui}}, \bibinfo {author} {\bibfnamefont {J.~P.}\ \bibnamefont {Ostriker}},
  \bibinfo {author} {\bibfnamefont {S.}~\bibnamefont {Tremaine}}, \ and\
  \bibinfo {author} {\bibfnamefont {E.}~\bibnamefont {Witten}},\ }\href
  {\doibase 10.1103/PhysRevD.95.043541} {\bibfield  {journal} {\bibinfo
  {journal} {Phys. Rev.}\ }\textbf {\bibinfo {volume} {D95}},\ \bibinfo {pages}
  {043541} (\bibinfo {year} {2017})},\ \Eprint
  {http://arxiv.org/abs/1610.08297} {arXiv:1610.08297 [astro-ph.CO]}
  \BibitemShut {NoStop}%
\bibitem [{\citenamefont {Berezhiani}\ and\ \citenamefont
  {Khoury}(2015)}]{Berezhiani:2015bqa}%
  \BibitemOpen
  \bibfield  {author} {\bibinfo {author} {\bibfnamefont {L.}~\bibnamefont
  {Berezhiani}}\ and\ \bibinfo {author} {\bibfnamefont {J.}~\bibnamefont
  {Khoury}},\ }\href {\doibase 10.1103/PhysRevD.92.103510} {\bibfield
  {journal} {\bibinfo  {journal} {Phys. Rev.}\ }\textbf {\bibinfo {volume}
  {D92}},\ \bibinfo {pages} {103510} (\bibinfo {year} {2015})},\ \Eprint
  {http://arxiv.org/abs/1507.01019} {arXiv:1507.01019 [astro-ph.CO]}
  \BibitemShut {NoStop}%
\bibitem [{\citenamefont {Ferreira}\ \emph {et~al.}(2018)\citenamefont
  {Ferreira}, \citenamefont {Franzmann}, \citenamefont {Khoury},\ and\
  \citenamefont {Brandenberger}}]{Ferreira:2018wup}%
  \BibitemOpen
  \bibfield  {author} {\bibinfo {author} {\bibfnamefont {E.~G.~M.}\
  \bibnamefont {Ferreira}}, \bibinfo {author} {\bibfnamefont {G.}~\bibnamefont
  {Franzmann}}, \bibinfo {author} {\bibfnamefont {J.}~\bibnamefont {Khoury}}, \
  and\ \bibinfo {author} {\bibfnamefont {R.}~\bibnamefont {Brandenberger}},\
  }\href@noop {} {\  (\bibinfo {year} {2018})},\ \Eprint
  {http://arxiv.org/abs/1810.09474} {arXiv:1810.09474 [astro-ph.CO]}
  \BibitemShut {NoStop}%
\bibitem [{\citenamefont {Alexander}\ and\ \citenamefont
  {Cormack}(2017)}]{Alexander:2016glq}%
  \BibitemOpen
  \bibfield  {author} {\bibinfo {author} {\bibfnamefont {S.}~\bibnamefont
  {Alexander}}\ and\ \bibinfo {author} {\bibfnamefont {S.}~\bibnamefont
  {Cormack}},\ }\href {\doibase 10.1088/1475-7516/2017/04/005} {\bibfield
  {journal} {\bibinfo  {journal} {JCAP}\ }\textbf {\bibinfo {volume} {1704}},\
  \bibinfo {pages} {005} (\bibinfo {year} {2017})},\ \Eprint
  {http://arxiv.org/abs/1607.08621} {arXiv:1607.08621 [astro-ph.CO]}
  \BibitemShut {NoStop}%
\bibitem [{\citenamefont {Alexander}\ \emph {et~al.}(2018)\citenamefont
  {Alexander}, \citenamefont {McDonough},\ and\ \citenamefont
  {Spergel}}]{Alexander:2018fjp}%
  \BibitemOpen
  \bibfield  {author} {\bibinfo {author} {\bibfnamefont {S.}~\bibnamefont
  {Alexander}}, \bibinfo {author} {\bibfnamefont {E.}~\bibnamefont
  {McDonough}}, \ and\ \bibinfo {author} {\bibfnamefont {D.~N.}\ \bibnamefont
  {Spergel}},\ }\href {\doibase 10.1088/1475-7516/2018/05/003} {\bibfield
  {journal} {\bibinfo  {journal} {JCAP}\ }\textbf {\bibinfo {volume} {1805}},\
  \bibinfo {pages} {003} (\bibinfo {year} {2018})},\ \Eprint
  {http://arxiv.org/abs/1801.07255} {arXiv:1801.07255 [hep-th]} \BibitemShut
  {NoStop}%
\bibitem [{\citenamefont {Preskill}\ \emph {et~al.}(1983)\citenamefont
  {Preskill}, \citenamefont {Wise},\ and\ \citenamefont
  {Wilczek}}]{Preskill:1982cy}%
  \BibitemOpen
  \bibfield  {author} {\bibinfo {author} {\bibfnamefont {J.}~\bibnamefont
  {Preskill}}, \bibinfo {author} {\bibfnamefont {M.~B.}\ \bibnamefont {Wise}},
  \ and\ \bibinfo {author} {\bibfnamefont {F.}~\bibnamefont {Wilczek}},\ }\href
  {\doibase 10.1016/0370-2693(83)90637-8} {\bibfield  {journal} {\bibinfo
  {journal} {Phys. Lett.}\ }\textbf {\bibinfo {volume} {B120}},\ \bibinfo
  {pages} {127} (\bibinfo {year} {1983})},\ \bibinfo {note}
  {[,URL(1982)]}\BibitemShut {NoStop}%
CITATION = PHLTA,B120,127
\bibitem [{\citenamefont {Abbott}\ and\ \citenamefont
  {Sikivie}(1983)}]{Abbott:1982af}%
  \BibitemOpen
  \bibfield  {author} {\bibinfo {author} {\bibfnamefont {L.~F.}\ \bibnamefont
  {Abbott}}\ and\ \bibinfo {author} {\bibfnamefont {P.}~\bibnamefont
  {Sikivie}},\ }\href {\doibase 10.1016/0370-2693(83)90638-X} {\bibfield
  {journal} {\bibinfo  {journal} {Phys. Lett.}\ }\textbf {\bibinfo {volume}
  {B120}},\ \bibinfo {pages} {133} (\bibinfo {year} {1983})},\ \bibinfo {note}
  {[,URL(1982)]}\BibitemShut {NoStop}%
\bibitem [{\citenamefont {Dine}\ and\ \citenamefont
  {Fischler}(1983)}]{Dine:1982ah}%
  \BibitemOpen
  \bibfield  {author} {\bibinfo {author} {\bibfnamefont {M.}~\bibnamefont
  {Dine}}\ and\ \bibinfo {author} {\bibfnamefont {W.}~\bibnamefont
  {Fischler}},\ }\href {\doibase 10.1016/0370-2693(83)90639-1} {\bibfield
  {journal} {\bibinfo  {journal} {Phys. Lett.}\ }\textbf {\bibinfo {volume}
  {B120}},\ \bibinfo {pages} {137} (\bibinfo {year} {1983})},\ \bibinfo {note}
  {[,URL(1982)]}\BibitemShut {NoStop}%
\bibitem [{\citenamefont {Schmitt}(2015)}]{Schmitt:2014eka}%
  \BibitemOpen
  \bibfield  {author} {\bibinfo {author} {\bibfnamefont {A.}~\bibnamefont
  {Schmitt}},\ }\href {\doibase 10.1007/978-3-319-07947-9} {\bibfield
  {journal} {\bibinfo  {journal} {Lect. Notes Phys.}\ }\textbf {\bibinfo
  {volume} {888}},\ \bibinfo {pages} {pp.1} (\bibinfo {year} {2015})},\ \Eprint
  {http://arxiv.org/abs/1404.1284} {arXiv:1404.1284 [hep-ph]} \BibitemShut
  {NoStop}%
\bibitem [{\citenamefont {Berezhiani}\ and\ \citenamefont
  {Khoury}(2016)}]{Berezhiani:2015pia}%
  \BibitemOpen
  \bibfield  {author} {\bibinfo {author} {\bibfnamefont {L.}~\bibnamefont
  {Berezhiani}}\ and\ \bibinfo {author} {\bibfnamefont {J.}~\bibnamefont
  {Khoury}},\ }\href {\doibase 10.1016/j.physletb.2015.12.054} {\bibfield
  {journal} {\bibinfo  {journal} {Phys. Lett.}\ }\textbf {\bibinfo {volume}
  {B753}},\ \bibinfo {pages} {639} (\bibinfo {year} {2016})},\ \Eprint
  {http://arxiv.org/abs/1506.07877} {arXiv:1506.07877 [astro-ph.CO]}
  \BibitemShut {NoStop}%
\bibitem [{\citenamefont {{T. Rindler-Daller, P.R. Shapiro}}(2012)}]{vort}%
  \BibitemOpen
  \bibfield  {author} {\bibinfo {author} {\bibnamefont {{T. Rindler-Daller,
  P.R. Shapiro}}},\ }\href@noop {} {\bibfield  {journal} {\bibinfo  {journal}
  {MNRAS}\ }\textbf {\bibinfo {volume} {422}},\ \bibinfo {pages} {135}
  (\bibinfo {year} {2012})},\ \bibinfo {note}
  {\href{https://arxiv.org/abs/1106.1256v4}{arXiv}}\BibitemShut {NoStop}%
\bibitem [{\citenamefont {Brandenberger}(1994)}]{Brandenberger:1993by}%
  \BibitemOpen
  \bibfield  {author} {\bibinfo {author} {\bibfnamefont {R.~H.}\ \bibnamefont
  {Brandenberger}},\ }\href {\doibase 10.1142/S0217751X9400090X} {\bibfield
  {journal} {\bibinfo  {journal} {Int. J. Mod. Phys.}\ }\textbf {\bibinfo
  {volume} {A9}},\ \bibinfo {pages} {2117} (\bibinfo {year} {1994})},\ \Eprint
  {http://arxiv.org/abs/astro-ph/9310041} {arXiv:astro-ph/9310041 [astro-ph]}
  \BibitemShut {NoStop}%
\bibitem [{\citenamefont {Brandenberger}(2014)}]{Brandenberger:2013tr}%
  \BibitemOpen
  \bibfield  {author} {\bibinfo {author} {\bibfnamefont {R.~H.}\ \bibnamefont
  {Brandenberger}},\ }\bibfield  {booktitle} {\emph {\bibinfo {booktitle}
  {{Proceedings, 9th International Symposium on Cosmology and Particle
  Astrophysics (CosPA 2012): Taipei, Taiwan, November 13-17, 2012}}},\ }\href
  {\doibase 10.1016/j.nuclphysbps.2013.10.064} {\bibfield  {journal} {\bibinfo
  {journal} {Nucl. Phys. Proc. Suppl.}\ }\textbf {\bibinfo {volume}
  {246-247}},\ \bibinfo {pages} {45} (\bibinfo {year} {2014})},\ \Eprint
  {http://arxiv.org/abs/1301.2856} {arXiv:1301.2856 [astro-ph.CO]} \BibitemShut
  {NoStop}%
\bibitem [{\citenamefont {{S. Mao and P. Schneider}}(1998)}]{sub1}%
  \BibitemOpen
  \bibfield  {author} {\bibinfo {author} {\bibnamefont {{S. Mao and P.
  Schneider}}},\ }\href@noop {} {\bibfield  {journal} {\bibinfo  {journal}
  {MNRAS}\ }\textbf {\bibinfo {volume} {295}},\ \bibinfo {pages} {587}
  (\bibinfo {year} {1998})},\ \bibinfo {note}
  {\href{https://arxiv.org/abs/astro-ph/9707187}{arXiv}}\BibitemShut {NoStop}%
\bibitem [{\citenamefont {{J.W. Hsueh et al.}}(2017)}]{sub2}%
  \BibitemOpen
  \bibfield  {author} {\bibinfo {author} {\bibnamefont {{J.W. Hsueh et al.}}},\
  }\href@noop {} {\bibfield  {journal} {\bibinfo  {journal} {MNRAS}\ }\textbf
  {\bibinfo {volume} {469}},\ \bibinfo {pages} {3713} (\bibinfo {year}
  {2017})},\ \bibinfo {note}
  {\href{https://arxiv.org/abs/1701.06575}{arXiv}}\BibitemShut {NoStop}%
\bibitem [{\citenamefont {{N. Dalal and C.S. Kochanek}}(2002)}]{sub3}%
  \BibitemOpen
  \bibfield  {author} {\bibinfo {author} {\bibnamefont {{N. Dalal and C.S.
  Kochanek}}},\ }\href@noop {} {\bibfield  {journal} {\bibinfo  {journal}
  {ApJ}\ }\textbf {\bibinfo {volume} {572}},\ \bibinfo {pages} {25} (\bibinfo
  {year} {2002})},\ \bibinfo {note}
  {\href{https://arxiv.org/abs/astro-ph/0111456}{arXiv}}\BibitemShut {NoStop}%
\bibitem [{\citenamefont {{Y.D. Hezaveh et al.}}(2016)}]{alma}%
  \BibitemOpen
  \bibfield  {author} {\bibinfo {author} {\bibnamefont {{Y.D. Hezaveh et
  al.}}},\ }\href@noop {} {\bibfield  {journal} {\bibinfo  {journal} {ApJ}\
  }\textbf {\bibinfo {volume} {823}},\ \bibinfo {pages} {37} (\bibinfo {year}
  {2016})},\ \bibinfo {note}
  {\href{https://arxiv.org/abs/1601.01388}{arXiv}}\BibitemShut {NoStop}%
\bibitem [{\citenamefont {{S. Vegetti and L.V.E.
  Koopmans}}(2009{\natexlab{a}})}]{veg}%
  \BibitemOpen
  \bibfield  {author} {\bibinfo {author} {\bibnamefont {{S. Vegetti and L.V.E.
  Koopmans}}},\ }\href@noop {} {\bibfield  {journal} {\bibinfo  {journal}
  {MNRAS}\ }\textbf {\bibinfo {volume} {392}},\ \bibinfo {pages} {945}
  (\bibinfo {year} {2009}{\natexlab{a}})},\ \bibinfo {note}
  {\href{https://arxiv.org/abs/0805.0201}{arXiv}}\BibitemShut {NoStop}%
\bibitem [{\citenamefont {{L.V.E. Koopmans}}(2005)}]{koop}%
  \BibitemOpen
  \bibfield  {author} {\bibinfo {author} {\bibnamefont {{L.V.E. Koopmans}}},\
  }\href@noop {} {\bibfield  {journal} {\bibinfo  {journal} {MNRAS}\ }\textbf
  {\bibinfo {volume} {363}},\ \bibinfo {pages} {1136} (\bibinfo {year}
  {2005})},\ \bibinfo {note}
  {\href{https://academic.oup.com/mnras/article/363/4/1136/1044360}{Oxford
  Journals}}\BibitemShut {NoStop}%
\bibitem [{\citenamefont {{S. Vegetti and L.V.E.
  Koopmans}}(2009{\natexlab{b}})}]{veko}%
  \BibitemOpen
  \bibfield  {author} {\bibinfo {author} {\bibnamefont {{S. Vegetti and L.V.E.
  Koopmans}}},\ }\href@noop {} {\bibfield  {journal} {\bibinfo  {journal}
  {MNRAS}\ }\textbf {\bibinfo {volume} {400}},\ \bibinfo {pages} {1583}
  (\bibinfo {year} {2009}{\natexlab{b}})},\ \bibinfo {note}
  {\href{https://arxiv.org/abs/0903.4752}{arXiv}}\BibitemShut {NoStop}%
\bibitem [{\citenamefont {{T. Daylan et al.}}(2018)}]{pcat}%
  \BibitemOpen
  \bibfield  {author} {\bibinfo {author} {\bibnamefont {{T. Daylan et al.}}},\
  }\href@noop {} {\bibfield  {journal} {\bibinfo  {journal} {ApJ}\ }\textbf
  {\bibinfo {volume} {854}},\ \bibinfo {pages} {141} (\bibinfo {year}
  {2018})},\ \bibinfo {note}
  {\href{https://arxiv.org/abs/1706.06111}{arXiv}}\BibitemShut {NoStop}%
\bibitem [{\citenamefont {{S. Vegetti et al.}}(2010)}]{subs}%
  \BibitemOpen
  \bibfield  {author} {\bibinfo {author} {\bibnamefont {{S. Vegetti et al.}}},\
  }\href@noop {} {\bibfield  {journal} {\bibinfo  {journal} {MNRAS}\ }\textbf
  {\bibinfo {volume} {408}},\ \bibinfo {pages} {1969} (\bibinfo {year}
  {2010})},\ \bibinfo {note}
  {\href{https://arxiv.org/abs/0910.0760v2}{arXiv}}\BibitemShut {NoStop}%
\bibitem [{\citenamefont {Ntampaka}\ \emph {et~al.}(2019)\citenamefont
  {Ntampaka} \emph {et~al.}}]{Ntampaka:2019udw}%
  \BibitemOpen
  \bibfield  {author} {\bibinfo {author} {\bibfnamefont {M.}~\bibnamefont
  {Ntampaka}} \emph {et~al.},\ }\href@noop {} {\  (\bibinfo {year} {2019})},\
  \Eprint {http://arxiv.org/abs/1902.10159} {arXiv:1902.10159 [astro-ph.IM]}
  \BibitemShut {NoStop}%
\bibitem [{\citenamefont {Carleo}\ \emph {et~al.}(2019)\citenamefont {Carleo},
  \citenamefont {Cirac}, \citenamefont {Cranmer}, \citenamefont {Daudet},
  \citenamefont {Schuld}, \citenamefont {Tishby}, \citenamefont
  {Vogt-Maranto},\ and\ \citenamefont {Zdeborová}}]{Carleo:2019ptp}%
  \BibitemOpen
  \bibfield  {author} {\bibinfo {author} {\bibfnamefont {G.}~\bibnamefont
  {Carleo}}, \bibinfo {author} {\bibfnamefont {I.}~\bibnamefont {Cirac}},
  \bibinfo {author} {\bibfnamefont {K.}~\bibnamefont {Cranmer}}, \bibinfo
  {author} {\bibfnamefont {L.}~\bibnamefont {Daudet}}, \bibinfo {author}
  {\bibfnamefont {M.}~\bibnamefont {Schuld}}, \bibinfo {author} {\bibfnamefont
  {N.}~\bibnamefont {Tishby}}, \bibinfo {author} {\bibfnamefont
  {L.}~\bibnamefont {Vogt-Maranto}}, \ and\ \bibinfo {author} {\bibfnamefont
  {L.}~\bibnamefont {Zdeborová}},\ }\href@noop {} {\  (\bibinfo {year}
  {2019})},\ \Eprint {http://arxiv.org/abs/1903.10563} {arXiv:1903.10563
  [physics.comp-ph]} \BibitemShut {NoStop}%
\bibitem [{\citenamefont {Hezaveh}\ \emph {et~al.}(2017)\citenamefont
  {Hezaveh}, \citenamefont {Perreault~Levasseur},\ and\ \citenamefont
  {Marshall}}]{Hezaveh:2017sht}%
  \BibitemOpen
  \bibfield  {author} {\bibinfo {author} {\bibfnamefont {Y.~D.}\ \bibnamefont
  {Hezaveh}}, \bibinfo {author} {\bibfnamefont {L.}~\bibnamefont
  {Perreault~Levasseur}}, \ and\ \bibinfo {author} {\bibfnamefont {P.~J.}\
  \bibnamefont {Marshall}},\ }\href {\doibase 10.1038/nature23463} {\bibfield
  {journal} {\bibinfo  {journal} {Nature}\ }\textbf {\bibinfo {volume} {548}},\
  \bibinfo {pages} {555} (\bibinfo {year} {2017})},\ \Eprint
  {http://arxiv.org/abs/1708.08842} {arXiv:1708.08842 [astro-ph.IM]}
  \BibitemShut {NoStop}%
\bibitem [{\citenamefont {Perreault~Levasseur}\ \emph
  {et~al.}(2017)\citenamefont {Perreault~Levasseur}, \citenamefont {Hezaveh},\
  and\ \citenamefont {Wechsler}}]{PerreaultLevasseur:2017ltk}%
  \BibitemOpen
  \bibfield  {author} {\bibinfo {author} {\bibfnamefont {L.}~\bibnamefont
  {Perreault~Levasseur}}, \bibinfo {author} {\bibfnamefont {Y.~D.}\
  \bibnamefont {Hezaveh}}, \ and\ \bibinfo {author} {\bibfnamefont {R.~H.}\
  \bibnamefont {Wechsler}},\ }\href {\doibase 10.3847/2041-8213/aa9704}
  {\bibfield  {journal} {\bibinfo  {journal} {Astrophys. J.}\ }\textbf
  {\bibinfo {volume} {850}},\ \bibinfo {pages} {L7} (\bibinfo {year} {2017})},\
  \Eprint {http://arxiv.org/abs/1708.08843} {arXiv:1708.08843 [astro-ph.CO]}
  \BibitemShut {NoStop}%
\bibitem [{\citenamefont {Morningstar}\ \emph {et~al.}(2018)\citenamefont
  {Morningstar}, \citenamefont {Hezaveh}, \citenamefont {Perreault~Levasseur},
  \citenamefont {Blandford}, \citenamefont {Marshall}, \citenamefont {Putzky},\
  and\ \citenamefont {Wechsler}}]{Morningstar:2018ase}%
  \BibitemOpen
  \bibfield  {author} {\bibinfo {author} {\bibfnamefont {W.~R.}\ \bibnamefont
  {Morningstar}}, \bibinfo {author} {\bibfnamefont {Y.~D.}\ \bibnamefont
  {Hezaveh}}, \bibinfo {author} {\bibfnamefont {L.}~\bibnamefont
  {Perreault~Levasseur}}, \bibinfo {author} {\bibfnamefont {R.~D.}\
  \bibnamefont {Blandford}}, \bibinfo {author} {\bibfnamefont {P.~J.}\
  \bibnamefont {Marshall}}, \bibinfo {author} {\bibfnamefont {P.}~\bibnamefont
  {Putzky}}, \ and\ \bibinfo {author} {\bibfnamefont {R.~H.}\ \bibnamefont
  {Wechsler}},\ }\href@noop {} {\  (\bibinfo {year} {2018})},\ \Eprint
  {http://arxiv.org/abs/1808.00011} {arXiv:1808.00011 [astro-ph.IM]}
  \BibitemShut {NoStop}%
\bibitem [{\citenamefont {Morningstar}\ \emph {et~al.}(2019)\citenamefont
  {Morningstar}, \citenamefont {Perreault~Levasseur}, \citenamefont {Hezaveh},
  \citenamefont {Blandford}, \citenamefont {Marshall}, \citenamefont {Putzky},
  \citenamefont {Rueter}, \citenamefont {Wechsler},\ and\ \citenamefont
  {Welling}}]{Morningstar:2019szx}%
  \BibitemOpen
  \bibfield  {author} {\bibinfo {author} {\bibfnamefont {W.~R.}\ \bibnamefont
  {Morningstar}}, \bibinfo {author} {\bibfnamefont {L.}~\bibnamefont
  {Perreault~Levasseur}}, \bibinfo {author} {\bibfnamefont {Y.~D.}\
  \bibnamefont {Hezaveh}}, \bibinfo {author} {\bibfnamefont {R.}~\bibnamefont
  {Blandford}}, \bibinfo {author} {\bibfnamefont {P.}~\bibnamefont {Marshall}},
  \bibinfo {author} {\bibfnamefont {P.}~\bibnamefont {Putzky}}, \bibinfo
  {author} {\bibfnamefont {T.~D.}\ \bibnamefont {Rueter}}, \bibinfo {author}
  {\bibfnamefont {R.}~\bibnamefont {Wechsler}}, \ and\ \bibinfo {author}
  {\bibfnamefont {M.}~\bibnamefont {Welling}},\ }\href@noop {} {\  (\bibinfo
  {year} {2019})},\ \Eprint {http://arxiv.org/abs/1901.01359} {arXiv:1901.01359
  [astro-ph.IM]} \BibitemShut {NoStop}%
\bibitem [{\citenamefont {Brehmer}\ \emph {et~al.}(2019)\citenamefont
  {Brehmer}, \citenamefont {Mishra-Sharma}, \citenamefont {Hermans},
  \citenamefont {Louppe},\ and\ \citenamefont {Cranmer}}]{Brehmer:2019jyt}%
  \BibitemOpen
  \bibfield  {author} {\bibinfo {author} {\bibfnamefont {J.}~\bibnamefont
  {Brehmer}}, \bibinfo {author} {\bibfnamefont {S.}~\bibnamefont
  {Mishra-Sharma}}, \bibinfo {author} {\bibfnamefont {J.}~\bibnamefont
  {Hermans}}, \bibinfo {author} {\bibfnamefont {G.}~\bibnamefont {Louppe}}, \
  and\ \bibinfo {author} {\bibfnamefont {K.}~\bibnamefont {Cranmer}},\
  }\href@noop {} {\  (\bibinfo {year} {2019})},\ \Eprint
  {http://arxiv.org/abs/1909.02005} {arXiv:1909.02005 [astro-ph.CO]}
  \BibitemShut {NoStop}%
\bibitem [{\citenamefont {Diaz~Rivero}\ \emph {et~al.}(2018)\citenamefont
  {Diaz~Rivero}, \citenamefont {Cyr-Racine},\ and\ \citenamefont
  {Dvorkin}}]{Rivero:2017mao}%
  \BibitemOpen
  \bibfield  {author} {\bibinfo {author} {\bibfnamefont {A.}~\bibnamefont
  {Diaz~Rivero}}, \bibinfo {author} {\bibfnamefont {F.-Y.}\ \bibnamefont
  {Cyr-Racine}}, \ and\ \bibinfo {author} {\bibfnamefont {C.}~\bibnamefont
  {Dvorkin}},\ }\href {\doibase 10.1103/PhysRevD.97.023001} {\bibfield
  {journal} {\bibinfo  {journal} {Phys. Rev.}\ }\textbf {\bibinfo {volume}
  {D97}},\ \bibinfo {pages} {023001} (\bibinfo {year} {2018})},\ \Eprint
  {http://arxiv.org/abs/1707.04590} {arXiv:1707.04590 [astro-ph.CO]}
  \BibitemShut {NoStop}%
\bibitem [{\citenamefont {Cyr-Racine}\ \emph {et~al.}(2016)\citenamefont
  {Cyr-Racine}, \citenamefont {Moustakas}, \citenamefont {Keeton},
  \citenamefont {Sigurdson},\ and\ \citenamefont
  {Gilman}}]{Cyr-Racine:2015jwa}%
  \BibitemOpen
  \bibfield  {author} {\bibinfo {author} {\bibfnamefont {F.-Y.}\ \bibnamefont
  {Cyr-Racine}}, \bibinfo {author} {\bibfnamefont {L.~A.}\ \bibnamefont
  {Moustakas}}, \bibinfo {author} {\bibfnamefont {C.~R.}\ \bibnamefont
  {Keeton}}, \bibinfo {author} {\bibfnamefont {K.}~\bibnamefont {Sigurdson}}, \
  and\ \bibinfo {author} {\bibfnamefont {D.~A.}\ \bibnamefont {Gilman}},\
  }\href {\doibase 10.1103/PhysRevD.94.043505} {\bibfield  {journal} {\bibinfo
  {journal} {Phys. Rev.}\ }\textbf {\bibinfo {volume} {D94}},\ \bibinfo {pages}
  {043505} (\bibinfo {year} {2016})},\ \Eprint
  {http://arxiv.org/abs/1506.01724} {arXiv:1506.01724 [astro-ph.CO]}
  \BibitemShut {NoStop}%
\bibitem [{\citenamefont {Cyr-Racine}\ \emph {et~al.}(2018)\citenamefont
  {Cyr-Racine}, \citenamefont {Keeton},\ and\ \citenamefont
  {Moustakas}}]{Cyr-Racine:2018htu}%
  \BibitemOpen
  \bibfield  {author} {\bibinfo {author} {\bibfnamefont {F.-Y.}\ \bibnamefont
  {Cyr-Racine}}, \bibinfo {author} {\bibfnamefont {C.~R.}\ \bibnamefont
  {Keeton}}, \ and\ \bibinfo {author} {\bibfnamefont {L.~A.}\ \bibnamefont
  {Moustakas}},\ }\href@noop {} {\  (\bibinfo {year} {2018})},\ \Eprint
  {http://arxiv.org/abs/1806.07897} {arXiv:1806.07897 [astro-ph.CO]}
  \BibitemShut {NoStop}%
\bibitem [{\citenamefont {Díaz~Rivero}\ \emph {et~al.}(2018)\citenamefont
  {Díaz~Rivero}, \citenamefont {Dvorkin}, \citenamefont {Cyr-Racine},
  \citenamefont {Zavala},\ and\ \citenamefont {Vogelsberger}}]{Rivero:2018bcd}%
  \BibitemOpen
  \bibfield  {author} {\bibinfo {author} {\bibfnamefont {A.}~\bibnamefont
  {Díaz~Rivero}}, \bibinfo {author} {\bibfnamefont {C.}~\bibnamefont
  {Dvorkin}}, \bibinfo {author} {\bibfnamefont {F.-Y.}\ \bibnamefont
  {Cyr-Racine}}, \bibinfo {author} {\bibfnamefont {J.}~\bibnamefont {Zavala}},
  \ and\ \bibinfo {author} {\bibfnamefont {M.}~\bibnamefont {Vogelsberger}},\
  }\href {\doibase 10.1103/PhysRevD.98.103517} {\bibfield  {journal} {\bibinfo
  {journal} {Phys. Rev.}\ }\textbf {\bibinfo {volume} {D98}},\ \bibinfo {pages}
  {103517} (\bibinfo {year} {2018})},\ \Eprint
  {http://arxiv.org/abs/1809.00004} {arXiv:1809.00004 [astro-ph.CO]}
  \BibitemShut {NoStop}%
\bibitem [{\citenamefont {Brennan}\ \emph {et~al.}(2018)\citenamefont
  {Brennan}, \citenamefont {Benson}, \citenamefont {Cyr-Racine}, \citenamefont
  {Keeton}, \citenamefont {Moustakas},\ and\ \citenamefont
  {Pullen}}]{Brennan:2018jhq}%
  \BibitemOpen
  \bibfield  {author} {\bibinfo {author} {\bibfnamefont {S.}~\bibnamefont
  {Brennan}}, \bibinfo {author} {\bibfnamefont {A.~J.}\ \bibnamefont {Benson}},
  \bibinfo {author} {\bibfnamefont {F.-Y.}\ \bibnamefont {Cyr-Racine}},
  \bibinfo {author} {\bibfnamefont {C.~R.}\ \bibnamefont {Keeton}}, \bibinfo
  {author} {\bibfnamefont {L.~A.}\ \bibnamefont {Moustakas}}, \ and\ \bibinfo
  {author} {\bibfnamefont {A.~R.}\ \bibnamefont {Pullen}},\ }\href@noop {} {\
  (\bibinfo {year} {2018})},\ \Eprint {http://arxiv.org/abs/1808.03501}
  {arXiv:1808.03501 [astro-ph.GA]} \BibitemShut {NoStop}%
\bibitem [{\citenamefont {Kauffmann}\ \emph {et~al.}(1993)\citenamefont
  {Kauffmann}, \citenamefont {White},\ and\ \citenamefont
  {Guiderdoni}}]{Kauffmann:1993gv}%
  \BibitemOpen
  \bibfield  {author} {\bibinfo {author} {\bibfnamefont {G.}~\bibnamefont
  {Kauffmann}}, \bibinfo {author} {\bibfnamefont {S.~D.~M.}\ \bibnamefont
  {White}}, \ and\ \bibinfo {author} {\bibfnamefont {B.}~\bibnamefont
  {Guiderdoni}},\ }\href@noop {} {\bibfield  {journal} {\bibinfo  {journal}
  {Mon. Not. Roy. Astron. Soc.}\ }\textbf {\bibinfo {volume} {264}},\ \bibinfo
  {pages} {201} (\bibinfo {year} {1993})}\BibitemShut {NoStop}%
\bibitem [{\citenamefont {{Planck Collaboration}}(2016)}]{planck}%
  \BibitemOpen
  \bibfield  {author} {\bibinfo {author} {\bibnamefont {{Planck
  Collaboration}}},\ }\href@noop {} {\bibfield  {journal} {\bibinfo  {journal}
  {A\&A}\ }\textbf {\bibinfo {volume} {594}},\ \bibinfo {pages} {63} (\bibinfo
  {year} {2016})},\ \bibinfo {note}
  {\href{https://arxiv.org/abs/1502.01589}{arXiv}}\BibitemShut {NoStop}%
\bibitem [{\citenamefont {{L. Anderson, E. Aubourg et al.}}(2014)}]{clust}%
  \BibitemOpen
  \bibfield  {author} {\bibinfo {author} {\bibnamefont {{L. Anderson, E.
  Aubourg et al.}}},\ }\href@noop {} {\bibfield  {journal} {\bibinfo  {journal}
  {MNRAS}\ }\textbf {\bibinfo {volume} {441}},\ \bibinfo {pages} {24} (\bibinfo
  {year} {2014})},\ \bibinfo {note}
  {\href{https://academic.oup.com/mnras/article/441/1/24/978049}{MNRAS}}\BibitemShut
  {NoStop}%
\bibitem [{\citenamefont {{C. Heymans, L. van Waerbeke et al.}}(2012)}]{wlens}%
  \BibitemOpen
  \bibfield  {author} {\bibinfo {author} {\bibnamefont {{C. Heymans, L. van
  Waerbeke et al.}}},\ }\href@noop {} {\bibfield  {journal} {\bibinfo
  {journal} {MNRAS}\ }\textbf {\bibinfo {volume} {427}},\ \bibinfo {pages}
  {146} (\bibinfo {year} {2012})},\ \bibinfo {note}
  {\href{https://academic.oup.com/mnras/article/427/1/146/1027955}{MNRAS}}\BibitemShut
  {NoStop}%
\bibitem [{\citenamefont {{ J. S. Bullock and M. Boylan-Kolch}}(2017)}]{msp}%
  \BibitemOpen
  \bibfield  {author} {\bibinfo {author} {\bibnamefont {{ J. S. Bullock and M.
  Boylan-Kolch}}},\ }\href@noop {} {\bibfield  {journal} {\bibinfo  {journal}
  {ARA\&A}\ }\textbf {\bibinfo {volume} {55}},\ \bibinfo {pages} {343}
  (\bibinfo {year} {2017})},\ \bibinfo {note}
  {\href{https://arxiv.org/abs/1707.04256}{arXiv}}\BibitemShut {NoStop}%
\bibitem [{\citenamefont {Kim}\ \emph {et~al.}(2018)\citenamefont {Kim},
  \citenamefont {Peter},\ and\ \citenamefont {Hargis}}]{mspr}%
  \BibitemOpen
  \bibfield  {author} {\bibinfo {author} {\bibfnamefont {S.~Y.}\ \bibnamefont
  {Kim}}, \bibinfo {author} {\bibfnamefont {A.~H.~G.}\ \bibnamefont {Peter}}, \
  and\ \bibinfo {author} {\bibfnamefont {J.~R.}\ \bibnamefont {Hargis}},\
  }\href {\doibase 10.1103/PhysRevLett.121.211302} {\bibfield  {journal}
  {\bibinfo  {journal} {Phys. Rev. Lett.}\ }\textbf {\bibinfo {volume} {121}},\
  \bibinfo {pages} {211302} (\bibinfo {year} {2018})},\ \Eprint
  {http://arxiv.org/abs/1711.06267} {arXiv:1711.06267 [astro-ph.CO]}
  \BibitemShut {NoStop}%
\bibitem [{\citenamefont {{P. Bode, J. P. Ostriker and N. Turok
  }}(2001)}]{wdm1}%
  \BibitemOpen
  \bibfield  {author} {\bibinfo {author} {\bibnamefont {{P. Bode, J. P.
  Ostriker and N. Turok }}},\ }\href@noop {} {\bibfield  {journal} {\bibinfo
  {journal} {ApJ}\ }\textbf {\bibinfo {volume} {556}},\ \bibinfo {pages} {93}
  (\bibinfo {year} {2001})},\ \bibinfo {note}
  {\href{https://arxiv.org/abs/astro-ph/0010389}{arXiv}}\BibitemShut {NoStop}%
\bibitem [{\citenamefont {{K. Abazajian}}(2006)}]{wdm2}%
  \BibitemOpen
  \bibfield  {author} {\bibinfo {author} {\bibnamefont {{K. Abazajian}}},\
  }\href@noop {} {\bibfield  {journal} {\bibinfo  {journal} {Phys. Rev. D.}\
  }\textbf {\bibinfo {volume} {73}},\ \bibinfo {pages} {063513} (\bibinfo
  {year} {2006})},\ \bibinfo {note}
  {\href{https://arxiv.org/abs/astro-ph/0512631}{arXiv}}\BibitemShut {NoStop}%
\bibitem [{\citenamefont {{D. N. Spergel and P. J. Steinhardt}}(2000)}]{idm}%
  \BibitemOpen
  \bibfield  {author} {\bibinfo {author} {\bibnamefont {{D. N. Spergel and P.
  J. Steinhardt}}},\ }\href@noop {} {\bibfield  {journal} {\bibinfo  {journal}
  {Phys. Rev. Lett.}\ }\textbf {\bibinfo {volume} {84}},\ \bibinfo {pages}
  {3760} (\bibinfo {year} {2000})},\ \bibinfo {note}
  {\href{https://arxiv.org/abs/astro-ph/9909386}{arXiv}}\BibitemShut {NoStop}%
\bibitem [{\citenamefont {Narayan}\ and\ \citenamefont
  {Bartelmann}(1997)}]{Nar-Bart:1997lens}%
  \BibitemOpen
  \bibfield  {author} {\bibinfo {author} {\bibfnamefont {R.}~\bibnamefont
  {Narayan}}\ and\ \bibinfo {author} {\bibfnamefont {M.}~\bibnamefont
  {Bartelmann}},\ }\href@noop {} {\  (\bibinfo {year} {1997})},\ \Eprint
  {http://arxiv.org/abs/9606001} {arXiv:9606001 [astro-ph.CO]} \BibitemShut
  {NoStop}%
\bibitem [{\citenamefont {Peccei}\ and\ \citenamefont
  {Quinn}(1977)}]{Peccei:1977hh}%
  \BibitemOpen
  \bibfield  {author} {\bibinfo {author} {\bibfnamefont {R.~D.}\ \bibnamefont
  {Peccei}}\ and\ \bibinfo {author} {\bibfnamefont {H.~R.}\ \bibnamefont
  {Quinn}},\ }\href {\doibase 10.1103/PhysRevLett.38.1440} {\bibfield
  {journal} {\bibinfo  {journal} {Phys. Rev. Lett.}\ }\textbf {\bibinfo
  {volume} {38}},\ \bibinfo {pages} {1440} (\bibinfo {year} {1977})},\ \bibinfo
  {note} {[,328(1977)]}\BibitemShut {NoStop}%
\bibitem [{\citenamefont {Wilczek}(1978)}]{Wilczek:1977pj}%
  \BibitemOpen
  \bibfield  {author} {\bibinfo {author} {\bibfnamefont {F.}~\bibnamefont
  {Wilczek}},\ }\href {\doibase 10.1103/PhysRevLett.40.279} {\bibfield
  {journal} {\bibinfo  {journal} {Phys. Rev. Lett.}\ }\textbf {\bibinfo
  {volume} {40}},\ \bibinfo {pages} {279} (\bibinfo {year} {1978})}\BibitemShut
  {NoStop}%
\bibitem [{\citenamefont {Weinberg}(1978)}]{Weinberg:1977ma}%
  \BibitemOpen
  \bibfield  {author} {\bibinfo {author} {\bibfnamefont {S.}~\bibnamefont
  {Weinberg}},\ }\href {\doibase 10.1103/PhysRevLett.40.223} {\bibfield
  {journal} {\bibinfo  {journal} {Phys. Rev. Lett.}\ }\textbf {\bibinfo
  {volume} {40}},\ \bibinfo {pages} {223} (\bibinfo {year} {1978})}\BibitemShut
  {NoStop}%
\bibitem [{\citenamefont {{Baym}}\ \emph {et~al.}(1969)\citenamefont {{Baym}},
  \citenamefont {{Pethick}},\ and\ \citenamefont
  {{Pines}}}]{1969Natur.224673B}%
  \BibitemOpen
  \bibfield  {author} {\bibinfo {author} {\bibfnamefont {G.}~\bibnamefont
  {{Baym}}}, \bibinfo {author} {\bibfnamefont {C.}~\bibnamefont {{Pethick}}}, \
  and\ \bibinfo {author} {\bibfnamefont {D.}~\bibnamefont {{Pines}}},\ }\href
  {\doibase 10.1038/224673a0} {\bibfield  {journal} {\bibinfo  {journal}
  {\nat}\ }\textbf {\bibinfo {volume} {224}},\ \bibinfo {pages} {673} (\bibinfo
  {year} {1969})}\BibitemShut {NoStop}%
\bibitem [{\citenamefont {Alford}\ \emph {et~al.}(1998)\citenamefont {Alford},
  \citenamefont {Rajagopal},\ and\ \citenamefont {Wilczek}}]{Alford:1997zt}%
  \BibitemOpen
  \bibfield  {author} {\bibinfo {author} {\bibfnamefont {M.~G.}\ \bibnamefont
  {Alford}}, \bibinfo {author} {\bibfnamefont {K.}~\bibnamefont {Rajagopal}}, \
  and\ \bibinfo {author} {\bibfnamefont {F.}~\bibnamefont {Wilczek}},\ }\href
  {\doibase 10.1016/S0370-2693(98)00051-3} {\bibfield  {journal} {\bibinfo
  {journal} {Phys. Lett.}\ }\textbf {\bibinfo {volume} {B422}},\ \bibinfo
  {pages} {247} (\bibinfo {year} {1998})},\ \Eprint
  {http://arxiv.org/abs/hep-ph/9711395} {arXiv:hep-ph/9711395 [hep-ph]}
  \BibitemShut {NoStop}%
\bibitem [{\citenamefont {Alford}\ \emph {et~al.}(2008)\citenamefont {Alford},
  \citenamefont {Schmitt}, \citenamefont {Rajagopal},\ and\ \citenamefont
  {Schäfer}}]{Alford:2007xm}%
  \BibitemOpen
  \bibfield  {author} {\bibinfo {author} {\bibfnamefont {M.~G.}\ \bibnamefont
  {Alford}}, \bibinfo {author} {\bibfnamefont {A.}~\bibnamefont {Schmitt}},
  \bibinfo {author} {\bibfnamefont {K.}~\bibnamefont {Rajagopal}}, \ and\
  \bibinfo {author} {\bibfnamefont {T.}~\bibnamefont {Schäfer}},\ }\href
  {\doibase 10.1103/RevModPhys.80.1455} {\bibfield  {journal} {\bibinfo
  {journal} {Rev. Mod. Phys.}\ }\textbf {\bibinfo {volume} {80}},\ \bibinfo
  {pages} {1455} (\bibinfo {year} {2008})},\ \Eprint
  {http://arxiv.org/abs/0709.4635} {arXiv:0709.4635 [hep-ph]} \BibitemShut
  {NoStop}%
\bibitem [{\citenamefont {Lombardo}\ and\ \citenamefont
  {Schulze}(2001)}]{Lombardo:2000ec}%
  \BibitemOpen
  \bibfield  {author} {\bibinfo {author} {\bibfnamefont {U.}~\bibnamefont
  {Lombardo}}\ and\ \bibinfo {author} {\bibfnamefont {H.~J.}\ \bibnamefont
  {Schulze}},\ }\bibfield  {booktitle} {\emph {\bibinfo {booktitle}
  {{Proceedings, ECT International Workshop on Physics of Neutron Star
  Interiors(NSI00): Trento, Italy, June 19-July 6, 2000}}},\ }\href@noop {}
  {\bibfield  {journal} {\bibinfo  {journal} {Lect. Notes Phys.}\ }\textbf
  {\bibinfo {volume} {578}},\ \bibinfo {pages} {30} (\bibinfo {year} {2001})},\
  \bibinfo {note} {[,30(2000)]},\ \Eprint
  {http://arxiv.org/abs/astro-ph/0012209} {arXiv:astro-ph/0012209 [astro-ph]}
  \BibitemShut {NoStop}%
\bibitem [{\citenamefont {Dean}\ and\ \citenamefont
  {Hjorth-Jensen}(2003)}]{Dean:2002zx}%
  \BibitemOpen
  \bibfield  {author} {\bibinfo {author} {\bibfnamefont {D.~J.}\ \bibnamefont
  {Dean}}\ and\ \bibinfo {author} {\bibfnamefont {M.}~\bibnamefont
  {Hjorth-Jensen}},\ }\href {\doibase 10.1103/RevModPhys.75.607} {\bibfield
  {journal} {\bibinfo  {journal} {Rev. Mod. Phys.}\ }\textbf {\bibinfo {volume}
  {75}},\ \bibinfo {pages} {607} (\bibinfo {year} {2003})},\ \Eprint
  {http://arxiv.org/abs/nucl-th/0210033} {arXiv:nucl-th/0210033 [nucl-th]}
  \BibitemShut {NoStop}%
\bibitem [{\citenamefont {Page}\ \emph {et~al.}(2013)\citenamefont {Page},
  \citenamefont {Lattimer}, \citenamefont {Prakash},\ and\ \citenamefont
  {Steiner}}]{Page:2013hxa}%
  \BibitemOpen
  \bibfield  {author} {\bibinfo {author} {\bibfnamefont {D.}~\bibnamefont
  {Page}}, \bibinfo {author} {\bibfnamefont {J.~M.}\ \bibnamefont {Lattimer}},
  \bibinfo {author} {\bibfnamefont {M.}~\bibnamefont {Prakash}}, \ and\
  \bibinfo {author} {\bibfnamefont {A.~W.}\ \bibnamefont {Steiner}},\
  }\href@noop {} {\  (\bibinfo {year} {2013})},\ \Eprint
  {http://arxiv.org/abs/1302.6626} {arXiv:1302.6626 [astro-ph.HE]} \BibitemShut
  {NoStop}%
\bibitem [{\citenamefont {Haskell}\ and\ \citenamefont
  {Sedrakian}(2018)}]{Haskell:2017lkl}%
  \BibitemOpen
  \bibfield  {author} {\bibinfo {author} {\bibfnamefont {B.}~\bibnamefont
  {Haskell}}\ and\ \bibinfo {author} {\bibfnamefont {A.}~\bibnamefont
  {Sedrakian}},\ }\href {\doibase 10.1007/978-3-319-97616-7_8} {\bibfield
  {journal} {\bibinfo  {journal} {Astrophys. Space Sci. Libr.}\ }\textbf
  {\bibinfo {volume} {457}},\ \bibinfo {pages} {401} (\bibinfo {year}
  {2018})},\ \Eprint {http://arxiv.org/abs/1709.10340} {arXiv:1709.10340
  [astro-ph.HE]} \BibitemShut {NoStop}%
\bibitem [{\citenamefont {Abbott}\ \emph {et~al.}(2017)\citenamefont {Abbott}
  \emph {et~al.}}]{TheLIGOScientific:2017qsa}%
  \BibitemOpen
  \bibfield  {author} {\bibinfo {author} {\bibfnamefont {B.~P.}\ \bibnamefont
  {Abbott}} \emph {et~al.} (\bibinfo {collaboration} {LIGO Scientific,
  Virgo}),\ }\href {\doibase 10.1103/PhysRevLett.119.161101} {\bibfield
  {journal} {\bibinfo  {journal} {Phys. Rev. Lett.}\ }\textbf {\bibinfo
  {volume} {119}},\ \bibinfo {pages} {161101} (\bibinfo {year} {2017})},\
  \Eprint {http://arxiv.org/abs/1710.05832} {arXiv:1710.05832 [gr-qc]}
  \BibitemShut {NoStop}%
\bibitem [{\citenamefont {{T. Rindler-Daller, P. R. Shapiro}}(2012)}]{sfdm}%
  \BibitemOpen
  \bibfield  {author} {\bibinfo {author} {\bibnamefont {{T. Rindler-Daller, P.
  R. Shapiro}}},\ }\href@noop {} {\bibfield  {journal} {\bibinfo  {journal}
  {MNRAS}\ }\textbf {\bibinfo {volume} {422}},\ \bibinfo {pages} {135}
  (\bibinfo {year} {2012})},\ \bibinfo {note}
  {\href{https://arxiv.org/abs/1106.1256}{arXiv:1106.1256}}\BibitemShut
  {NoStop}%
\bibitem [{\citenamefont {Banik}\ and\ \citenamefont
  {Sikivie}(2013)}]{Banik:2013rxa}%
  \BibitemOpen
  \bibfield  {author} {\bibinfo {author} {\bibfnamefont {N.}~\bibnamefont
  {Banik}}\ and\ \bibinfo {author} {\bibfnamefont {P.}~\bibnamefont
  {Sikivie}},\ }\href {\doibase 10.1103/PhysRevD.88.123517} {\bibfield
  {journal} {\bibinfo  {journal} {Phys. Rev.}\ }\textbf {\bibinfo {volume}
  {D88}},\ \bibinfo {pages} {123517} (\bibinfo {year} {2013})},\ \Eprint
  {http://arxiv.org/abs/1307.3547} {arXiv:1307.3547 [astro-ph.GA]} \BibitemShut
  {NoStop}%
\bibitem [{\citenamefont {Alexander}\ \emph {et~al.}(2019)\citenamefont
  {Alexander}, \citenamefont {Bramburger},\ and\ \citenamefont
  {McDonough}}]{Alexander:2019qsh}%
  \BibitemOpen
  \bibfield  {author} {\bibinfo {author} {\bibfnamefont {S.}~\bibnamefont
  {Alexander}}, \bibinfo {author} {\bibfnamefont {J.~J.}\ \bibnamefont
  {Bramburger}}, \ and\ \bibinfo {author} {\bibfnamefont {E.}~\bibnamefont
  {McDonough}},\ }\href@noop {} {\  (\bibinfo {year} {2019})},\ \Eprint
  {http://arxiv.org/abs/1901.03694} {arXiv:1901.03694 [astro-ph.CO]}
  \BibitemShut {NoStop}%
\bibitem [{\citenamefont {Sazhin}\ \emph {et~al.}(2007)\citenamefont {Sazhin},
  \citenamefont {Khovanskaya}, \citenamefont {Capaccioli}, \citenamefont
  {Longo}, \citenamefont {Paolillo}, \citenamefont {Covone}, \citenamefont
  {Grogin},\ and\ \citenamefont {Schreier}}]{Sazhin:2006kf}%
  \BibitemOpen
  \bibfield  {author} {\bibinfo {author} {\bibfnamefont {M.~V.}\ \bibnamefont
  {Sazhin}}, \bibinfo {author} {\bibfnamefont {O.~S.}\ \bibnamefont
  {Khovanskaya}}, \bibinfo {author} {\bibfnamefont {M.}~\bibnamefont
  {Capaccioli}}, \bibinfo {author} {\bibfnamefont {G.}~\bibnamefont {Longo}},
  \bibinfo {author} {\bibfnamefont {M.}~\bibnamefont {Paolillo}}, \bibinfo
  {author} {\bibfnamefont {G.}~\bibnamefont {Covone}}, \bibinfo {author}
  {\bibfnamefont {N.~A.}\ \bibnamefont {Grogin}}, \ and\ \bibinfo {author}
  {\bibfnamefont {E.~J.}\ \bibnamefont {Schreier}},\ }\href {\doibase
  10.1111/j.1365-2966.2007.11543.x} {\bibfield  {journal} {\bibinfo  {journal}
  {Mon. Not. Roy. Astron. Soc.}\ }\textbf {\bibinfo {volume} {376}},\ \bibinfo
  {pages} {1731} (\bibinfo {year} {2007})},\ \Eprint
  {http://arxiv.org/abs/astro-ph/0611744} {arXiv:astro-ph/0611744 [astro-ph]}
  \BibitemShut {NoStop}%
\bibitem [{\citenamefont {Gasparini}\ \emph {et~al.}(2008)\citenamefont
  {Gasparini}, \citenamefont {Marshall}, \citenamefont {Treu}, \citenamefont
  {Morganson},\ and\ \citenamefont {Dubath}}]{Gasparini:2007jj}%
  \BibitemOpen
  \bibfield  {author} {\bibinfo {author} {\bibfnamefont {M.~A.}\ \bibnamefont
  {Gasparini}}, \bibinfo {author} {\bibfnamefont {P.}~\bibnamefont {Marshall}},
  \bibinfo {author} {\bibfnamefont {T.}~\bibnamefont {Treu}}, \bibinfo {author}
  {\bibfnamefont {E.}~\bibnamefont {Morganson}}, \ and\ \bibinfo {author}
  {\bibfnamefont {F.}~\bibnamefont {Dubath}},\ }\href {\doibase
  10.1111/j.1365-2966.2007.12657.x} {\bibfield  {journal} {\bibinfo  {journal}
  {Mon. Not. Roy. Astron. Soc.}\ }\textbf {\bibinfo {volume} {385}},\ \bibinfo
  {pages} {1959} (\bibinfo {year} {2008})},\ \Eprint
  {http://arxiv.org/abs/0710.5544} {arXiv:0710.5544 [astro-ph]} \BibitemShut
  {NoStop}%
\bibitem [{\citenamefont {Morganson}\ \emph {et~al.}(2010)\citenamefont
  {Morganson}, \citenamefont {Marshall}, \citenamefont {Treu}, \citenamefont
  {Schrabback},\ and\ \citenamefont {Blandford}}]{Morganson:2009yk}%
  \BibitemOpen
  \bibfield  {author} {\bibinfo {author} {\bibfnamefont {E.}~\bibnamefont
  {Morganson}}, \bibinfo {author} {\bibfnamefont {P.}~\bibnamefont {Marshall}},
  \bibinfo {author} {\bibfnamefont {T.}~\bibnamefont {Treu}}, \bibinfo {author}
  {\bibfnamefont {T.}~\bibnamefont {Schrabback}}, \ and\ \bibinfo {author}
  {\bibfnamefont {R.~D.}\ \bibnamefont {Blandford}},\ }\href {\doibase
  10.1111/j.1365-2966.2010.16562.x} {\bibfield  {journal} {\bibinfo  {journal}
  {Mon. Not. Roy. Astron. Soc.}\ }\textbf {\bibinfo {volume} {406}},\ \bibinfo
  {pages} {2452} (\bibinfo {year} {2010})},\ \Eprint
  {http://arxiv.org/abs/0908.0602} {arXiv:0908.0602 [astro-ph.CO]} \BibitemShut
  {NoStop}%
\bibitem [{\citenamefont {{A. Verma, T. Collett et al.}}(2019)}]{lsstw}%
  \BibitemOpen
  \bibfield  {author} {\bibinfo {author} {\bibnamefont {{A. Verma, T. Collett
  et al.}}},\ }\href@noop {} {\  (\bibinfo {year} {2019})},\ \bibinfo {note}
  {\href{https://arxiv.org/abs/1902.05141}{arXiv}}\BibitemShut {NoStop}%
\bibitem [{\citenamefont {{J.W. Nightingale and S. Dye}}(2015)}]{night}%
  \BibitemOpen
  \bibfield  {author} {\bibinfo {author} {\bibnamefont {{J.W. Nightingale and
  S. Dye}}},\ }\href@noop {} {\bibfield  {journal} {\bibinfo  {journal}
  {MNRAS}\ }\textbf {\bibinfo {volume} {452}},\ \bibinfo {pages} {2940}
  (\bibinfo {year} {2015})},\ \bibinfo {note}
  {\href{https://arxiv.org/abs/1412.7436}{arXiv}}\BibitemShut {NoStop}%
\bibitem [{\citenamefont {{J.W. Nightingale, S. Dye, and R.J.
  Massey}}(2018)}]{night2}%
  \BibitemOpen
  \bibfield  {author} {\bibinfo {author} {\bibnamefont {{J.W. Nightingale, S.
  Dye, and R.J. Massey}}},\ }\href@noop {} {\bibfield  {journal} {\bibinfo
  {journal} {MNRAS}\ }\textbf {\bibinfo {volume} {478}},\ \bibinfo {pages}
  {4738} (\bibinfo {year} {2018})},\ \bibinfo {note}
  {\href{https://arxiv.org/abs/1708.07377}{arXiv}}\BibitemShut {NoStop}%
\bibitem [{\citenamefont {He}\ \emph {et~al.}(2015)\citenamefont {He},
  \citenamefont {Zhang}, \citenamefont {Ren},\ and\ \citenamefont
  {Sun}}]{DBLP:journals/corr/HeZRS15}%
  \BibitemOpen
  \bibfield  {author} {\bibinfo {author} {\bibfnamefont {K.}~\bibnamefont
  {He}}, \bibinfo {author} {\bibfnamefont {X.}~\bibnamefont {Zhang}}, \bibinfo
  {author} {\bibfnamefont {S.}~\bibnamefont {Ren}}, \ and\ \bibinfo {author}
  {\bibfnamefont {J.}~\bibnamefont {Sun}},\ }\href
  {http://arxiv.org/abs/1512.03385} {\bibfield  {journal} {\bibinfo  {journal}
  {CoRR}\ }\textbf {\bibinfo {volume} {abs/1512.03385}} (\bibinfo {year}
  {2015})},\ \Eprint {http://arxiv.org/abs/1512.03385} {arXiv:1512.03385}
  \BibitemShut {NoStop}%
\bibitem [{\citenamefont {Krizhevsky}\ \emph {et~al.}(2017)\citenamefont
  {Krizhevsky}, \citenamefont {Sutskever},\ and\ \citenamefont
  {Hinton}}]{Krizhevsky:2017:ICD:3098997.3065386}%
  \BibitemOpen
  \bibfield  {author} {\bibinfo {author} {\bibfnamefont {A.}~\bibnamefont
  {Krizhevsky}}, \bibinfo {author} {\bibfnamefont {I.}~\bibnamefont
  {Sutskever}}, \ and\ \bibinfo {author} {\bibfnamefont {G.~E.}\ \bibnamefont
  {Hinton}},\ }\href {\doibase 10.1145/3065386} {\bibfield  {journal} {\bibinfo
   {journal} {Commun. ACM}\ }\textbf {\bibinfo {volume} {60}},\ \bibinfo
  {pages} {84} (\bibinfo {year} {2017})}\BibitemShut {NoStop}%
\bibitem [{\citenamefont {Huang}\ \emph {et~al.}(2016)\citenamefont {Huang},
  \citenamefont {Liu},\ and\ \citenamefont
  {Weinberger}}]{DBLP:journals/corr/HuangLW16a}%
  \BibitemOpen
  \bibfield  {author} {\bibinfo {author} {\bibfnamefont {G.}~\bibnamefont
  {Huang}}, \bibinfo {author} {\bibfnamefont {Z.}~\bibnamefont {Liu}}, \ and\
  \bibinfo {author} {\bibfnamefont {K.~Q.}\ \bibnamefont {Weinberger}},\ }\href
  {http://arxiv.org/abs/1608.06993} {\bibfield  {journal} {\bibinfo  {journal}
  {CoRR}\ }\textbf {\bibinfo {volume} {abs/1608.06993}} (\bibinfo {year}
  {2016})},\ \Eprint {http://arxiv.org/abs/1608.06993} {arXiv:1608.06993}
  \BibitemShut {NoStop}%
\bibitem [{\citenamefont {{Simonyan}}\ and\ \citenamefont
  {{Zisserman}}(2014)}]{2014arXiv1409.1556S}%
  \BibitemOpen
  \bibfield  {author} {\bibinfo {author} {\bibfnamefont {K.}~\bibnamefont
  {{Simonyan}}}\ and\ \bibinfo {author} {\bibfnamefont {A.}~\bibnamefont
  {{Zisserman}}},\ }\href@noop {} {\bibfield  {journal} {\bibinfo  {journal}
  {arXiv e-prints}\ ,\ \bibinfo {eid} {arXiv:1409.1556}} (\bibinfo {year}
  {2014})},\ \Eprint {http://arxiv.org/abs/1409.1556} {arXiv:1409.1556 [cs.CV]}
  \BibitemShut {NoStop}%
\bibitem [{\citenamefont {Ciuca}\ \emph {et~al.}(2019)\citenamefont {Ciuca},
  \citenamefont {Hernández},\ and\ \citenamefont {Wolman}}]{Ciuca:2017gca}%
  \BibitemOpen
  \bibfield  {author} {\bibinfo {author} {\bibfnamefont {R.}~\bibnamefont
  {Ciuca}}, \bibinfo {author} {\bibfnamefont {O.~F.}\ \bibnamefont
  {Hernández}}, \ and\ \bibinfo {author} {\bibfnamefont {M.}~\bibnamefont
  {Wolman}},\ }\href {\doibase 10.1093/mnras/stz491} {\bibfield  {journal}
  {\bibinfo  {journal} {Mon. Not. Roy. Astron. Soc.}\ }\textbf {\bibinfo
  {volume} {485}},\ \bibinfo {pages} {1377} (\bibinfo {year} {2019})},\ \Eprint
  {http://arxiv.org/abs/1708.08878} {arXiv:1708.08878 [astro-ph.CO]}
  \BibitemShut {NoStop}%
\bibitem [{\citenamefont {Ciuca}\ and\ \citenamefont
  {Hernández}(2019)}]{Ciuca:2018tei}%
  \BibitemOpen
  \bibfield  {author} {\bibinfo {author} {\bibfnamefont {R.}~\bibnamefont
  {Ciuca}}\ and\ \bibinfo {author} {\bibfnamefont {O.~F.}\ \bibnamefont
  {Hernández}},\ }\href {\doibase 10.1093/mnras/sty3478} {\bibfield  {journal}
  {\bibinfo  {journal} {Mon. Not. Roy. Astron. Soc.}\ }\textbf {\bibinfo
  {volume} {483}},\ \bibinfo {pages} {5179} (\bibinfo {year} {2019})},\ \Eprint
  {http://arxiv.org/abs/1810.11889} {arXiv:1810.11889 [astro-ph.CO]}
  \BibitemShut {NoStop}%
\bibitem [{\citenamefont {Brandenberger}\ \emph {et~al.}(2010)\citenamefont
  {Brandenberger}, \citenamefont {Danos}, \citenamefont {Hernandez},\ and\
  \citenamefont {Holder}}]{Brandenberger:2010hn}%
  \BibitemOpen
  \bibfield  {author} {\bibinfo {author} {\bibfnamefont {R.~H.}\ \bibnamefont
  {Brandenberger}}, \bibinfo {author} {\bibfnamefont {R.~J.}\ \bibnamefont
  {Danos}}, \bibinfo {author} {\bibfnamefont {O.~F.}\ \bibnamefont
  {Hernandez}}, \ and\ \bibinfo {author} {\bibfnamefont {G.~P.}\ \bibnamefont
  {Holder}},\ }\href {\doibase 10.1088/1475-7516/2010/12/028} {\bibfield
  {journal} {\bibinfo  {journal} {JCAP}\ }\textbf {\bibinfo {volume} {1012}},\
  \bibinfo {pages} {028} (\bibinfo {year} {2010})},\ \Eprint
  {http://arxiv.org/abs/1006.2514} {arXiv:1006.2514 [astro-ph.CO]} \BibitemShut
  {NoStop}%
\bibitem [{\citenamefont {McDonough}\ and\ \citenamefont
  {Brandenberger}(2013)}]{McDonough:2011er}%
  \BibitemOpen
  \bibfield  {author} {\bibinfo {author} {\bibfnamefont {E.}~\bibnamefont
  {McDonough}}\ and\ \bibinfo {author} {\bibfnamefont {R.~H.}\ \bibnamefont
  {Brandenberger}},\ }\href {\doibase 10.1088/1475-7516/2013/02/045} {\bibfield
   {journal} {\bibinfo  {journal} {JCAP}\ }\textbf {\bibinfo {volume} {1302}},\
  \bibinfo {pages} {045} (\bibinfo {year} {2013})},\ \Eprint
  {http://arxiv.org/abs/1109.2627} {arXiv:1109.2627 [astro-ph.CO]} \BibitemShut
  {NoStop}%
\end{thebibliography}%

\end{document}